\documentclass[journal]{IEEEtran}
\IEEEoverridecommandlockouts

\usepackage{array}
\usepackage{bbm}
\usepackage[subrefformat=parens,labelformat=simple,farskip=0pt,justification=centering]{subfig}
\usepackage[labelsep=space,font=footnotesize]{caption}

\usepackage{balance}
\usepackage{graphicx}
\usepackage{authblk}
\usepackage{booktabs}
\usepackage{textcomp}   
\usepackage{stfloats}
\usepackage{url}
\usepackage{verbatim}
\usepackage{cite}
\usepackage{makecell}
\usepackage{multirow}
\usepackage{multicol}
\usepackage{comment}
\usepackage{enumitem}
\usepackage{threeparttable}
\usepackage{physunits}
\usepackage[round-pad=false, round-mode=places,round-precision=2,group-separator={,},output-decimal-marker={.}]{siunitx}
\usepackage{pgfplots}
\pgfplotsset{compat=newest}
\usepgfplotslibrary{groupplots}

\usepackage{etoolbox}
\makeatletter
\patchcmd{\@algocf@start}{-1.5em}{0pt}{}{} 
\newif\if@restonecol
\makeatother

\usepackage[noend,linesnumbered,ruled,vlined]{algorithm2e}
\usepackage{algpseudocode}
\usepackage{amsmath,amssymb,amsfonts}
\SetKwRepeat{Do}{do}{while}
\SetAlCapHSkip{0pt} 
\allowdisplaybreaks

\usepackage[bookmarks=false]{hyperref}
\hypersetup{
	colorlinks=true,
	linkcolor=blue,
	filecolor=black,      
	urlcolor=teal,
	citecolor=blue,
}

\usepackage{listings}
\lstdefinelanguage[riscv]{Assembler}%
{
morestring=[b]",
morekeywords = [1]{.alias,.align,.ascii,.asciiz,.byte,.data,.double,.end,.endb,%
  .endr,.ent,.err,.extern,.file,.float,.fmask,.frame,.globl,.half,.text,%
  .verstamp,.vreg,.word},%
morekeywords = [2]{lw,sw,lb,lbu, lh, lhu, sb, sh, la,li,lui, ori, or, sub, subi, add, addi, or, ori, xor, xori, sra, srai, srl, srli, slli, srl, sll, beq, bne,bge,blt, bltu, bgeu,j,jal,jalr, jr, j, or,addi,ecall,ret,load_patch, call, push, pop, jmp},%
morekeywords = [3]{x0,x1,x2,x3,x4,x5,x6,x7,x8,x9,x10,x11,x12,x13,x14,x15,x16,x17,x18,x19,x20,x21,ra,s0,s1,s2,a0,a1,x28,x29,x30,x31,sp,zero, a0,a1,a2,a3,a4,a5,a6,a7,sp,ra,t0,t1,t2,t3,t4,t5,t6,t7,s0,s1,s2,s3,s4,s5,s6,s7,s11},%
comment = [l]\#,%
keywordsprefix=.,%
sensitive=false,%
}[keywords,comments,strings]

\definecolor{customlightblue}{rgb}{0.8, 0.9, 1.0}

\lstdefinestyle{customriscv}{
  language = [riscv]Assembler,
  basicstyle = \footnotesize\ttfamily,
  stringstyle = \footnotesize\ttfamily,
  frame = single,
  rulesep = 5pt,
  numbers = none,
  framexleftmargin = \parindent,
  showspaces = false,
  showstringspaces = false,
  showtabs = false,
  rulecolor = \color{customlightblue},
  rulesepcolor = \color{customlightblue},
  tabsize=3,
  breaklines=false,
  keywordstyle = [1]\color{purple},%
  keywordstyle = [2]\color{blue},%
  keywordstyle = [3]\color{red},%
  stringstyle = \color{mauve},%
  escapeinside = {\%*}{*)},
  moredelim=[is][\bfseries\color{violet}]{;?}{?;},
  moredelim=[is][\bfseries\color{red}]{;!}{!;},
  moredelim=[is][\bfseries\color{green}]{;@}{@;},
  moredelim=[is][\bfseries]{[*}{*]},
  literate={à}{{\`a\ }}1 {é}{{\'e}}1 {è}{{\`e}}1 {ê}{{\^e}}1, %
  extendedchars=true
}

\usepackage{pifont} 
\usepackage{wasysym} 

\usepackage{colortbl} 
\usepackage[table]{xcolor} 
\definecolor{color_light}{HTML}{CBD7C3}
\definecolor{color_medium}{HTML}{ACBF9F}
\definecolor{color_dark}{HTML}{89AA78}

\newcommand{\cmark}{\ding{51}}%
\newcommand{\xmark}{\ding{55}}%
\usepackage{cleveref}
\Crefformat{figure}{Fig.~#2#1#3}                           
\Crefname{subfigure}{Fig.}{Figs.}
\Crefname{figure}{Fig.}{Figs.}
\Crefformat{table}{TABLE~#2#1#3}                           
\Crefname{table}{TABLE}{TABLEs}

\usepackage{xcolor}
\definecolor{SUATpurple}{RGB}{144,44,144}
\usepackage{tcolorbox}
\tcbuselibrary{skins,breakable}
{\endtcolorbox}
{\endtcolorbox}

\begin{document}

\title{Efficient Scheduling Exploration for Secure DNN Accelerators}
\newcommand{\design}{\textnormal{\emph{Seculor}}}

\author{\IEEEauthorblockN{Wei Xuan, Rongliang Fu, Zihao Xuan, Libo Shen, Lang Feng, Yuzhong Jiao, Tsung-Yi Ho~\IEEEmembership{Fellow,~IEEE}, and Zhongrui Wang}

  \thanks{This research was conducted by ACCESS – AI Chip Center for Emerging Smart Systems, supported by the InnoHK initiative of the Innovation and Technology Commission of the Hong Kong Special Administrative Region Government. It was also supported in part by Guangdong Provincial Key Laboratory of Computility Microelectronics (Grant No. 2024B1212010007) and Shenzhen Science and Technology Program (Grant No. ZDCY20250901112804006 and No. JCYJ20241206180301003). \textit{(Corresponding author: Zhongrui Wang.)}

    Wei Xuan, Zihao Xuan and Yuzhong Jiao are with the ACCESS – AI Chip Center for Emerging Smart Systems, InnoHK Centers, Hong Kong Science Park and the Hong Kong University of Science and Technology, Hong Kong, China (e-mail: \{weixuan, zihaoxuan, yuzhongjiao\}@ust.hk). Wei Xuan is also with Faculty of Computility Microelectronics, Shenzhen University of Advanced technology, Shenzhen, Guangdong, China.

    Rongliang Fu, Libo Shen and Tsung-Yi Ho are with The Chinese University of Hong Kong, Hong Kong 999077, China (e-mail: \{rlfu, lbshen24, tyho\}@cse.cuhk.edu.hk).

    Lang Feng is with Sun Yat-sen University Shenzhen Campus, Shenzhen, Guangdong, China (e-mail: flang1994@outlook.com).

    Zhongrui Wang is with Southern University of Science and Technology, Shenzhen, Guangdong, China (e-mail: wangzr@sustech.edu.cn).
  }
}
\IEEEaftertitletext{\vspace{-2\baselineskip}}

\maketitle


\begin{abstract}

Modern Deep Neural Networks (DNNs) are now deployed in domains such as autonomous driving, biomedical imaging, image recognition, and object detection, spurring the development of specialized hardware accelerators to meet stringent performance demands. In these settings, security is a first-order concern because model training is costly and both training and inference data may be sensitive. Although prior work has introduced protection mechanisms for DNN accelerators, two major inefficiencies remain. First, existing schemes often ignore tile-induced overlaps during computation, leading to redundant security operations and repeated data accesses. Second, they overlook the off-chip traffic caused by authentication metadata, which aggravates memory-bandwidth bottlenecks and degrades end-to-end performance.

To address these challenges, we propose {\design}, a scheduling framework for secure and efficient DNN accelerators that exploits the deterministic structure of inference workloads. {\design} introduces a fusion-aware chunking scheme to remove redundant security computations and data accesses caused by tile overlaps. Additionally, it adopts a two-level hierarchical authentication strategy that uses tile-based verification for partial sums and chunk-based verification for feature maps to reduce off-chip metadata accesses. {\design} integrates with existing DNN accelerator architectures without substantial hardware changes. Across representative workloads and hardware configurations, {\design} improves performance by $50.2\%$ on average and reduces the memory traffic of security metadata by $91.2\%$ compared with state-of-the-art secure DNN execution approaches.

\end{abstract}

\section{Introduction} \label{section:introduction}

\begin{figure}[!t]
    \centering
    \includegraphics[width=0.48\textwidth]{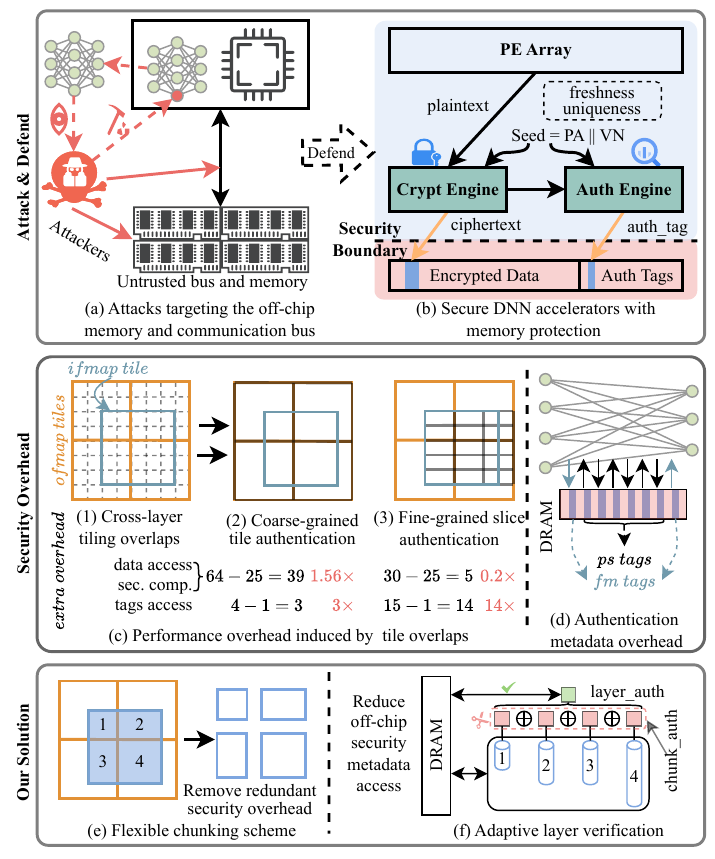}
    \caption{Overview of this paper.}
    \label{figures:introduction_overview}
    \vspace{-10pt}
\end{figure}

\IEEEPARstart{M}{odern} deep neural networks (DNNs)~\cite{krizhevsky2012imagenet,lecun2015deep,he2016deep} have become central to applications such as biomedical diagnosis~\cite{isensee2021nnu}, financial risk assessment~\cite{shi2025deep}, and autonomous driving~\cite{grigorescu2020survey}. Their widespread deployment in safety-critical applications renders security failures extremely costly, as any breach can lead to catastrophic financial losses and even endanger human lives. To meet stringent performance demands, dedicated DNN accelerators are widely deployed; however, they typically process sensitive model weights and intermediate feature maps in untrusted off-chip DRAM, leaving them directly exposed to physical adversaries. Hardware-level tampering and snooping attacks on DNN accelerators~\cite{lee2025secure} can expose proprietary models and sensitive data~\cite{yang2023huffduff}, or corrupt intermediate values and final predictions, as illustrated in \Cref{figures:introduction_overview}(a). In autonomous and embodied systems, such failures can directly translate into unsafe decisions. As a result, secure execution for DNN accelerators has become an important research problem for both academia and industry.

To address these threats, trusted execution environments (TEEs)~\cite{truong2021memory, johnson2021supporting, kim2020vessels} have emerged as a promising solution by leveraging authenticated encryption, which ensures both confidentiality and integrity during DNN execution, as shown in \Cref{figures:introduction_overview}(b). Specifically, existing TEE implementations~\cite{gassend2003caches, rogers2007using, saileshwar2018morphable} for DNN accelerators use \textit{encryption} to protect the confidentiality of model parameters and input data during computation and memory transfers, whereas \textit{authentication} ensures data integrity and defends against malicious tampering. To guarantee cryptographic security, the encryption seed utilizes physical addresses (PA) to achieve spatial uniqueness, whereas version numbers (VN), which track the count of data write operations, provide temporal uniqueness to mitigate replay attacks. Prior work~\cite{taassori2018vault,hua2022guardnn,hua2022mgx,lee2022tnpu, shrivastava2023securator} shows that such mechanisms can protect DNN execution with strong security guarantees, but their performance cost remains substantial. A key reason is that these mechanisms were originally designed for general-purpose workloads and do not account for the structured, tiled memory access patterns characteristic of DNN inference.
This cost is especially problematic for memory-bound accelerators, where extra cryptographic work and metadata movement can quickly dominate execution time.

However, current secure DNN acceleration approaches still incur two major sources of overhead. First, tile-induced overlap overhead arises because authentication blocks are poorly aligned with the tiled data blocks used during DNN computation. In the schematic of \Cref{figures:introduction_overview}(c), the current layer’s output feature map (ofmap) contains 64 elements with an 8×8 spatial dimension. Under the coarse-grained method, a single tile in the next layer's input feature map (ifmap) comprises 25 elements (5×5) and is associated with 4 authentication tags (authTags). In contrast, the fine-grained sliding method, which adopts a 1×2 authentication granularity, requires 30 element reads including 5 redundant ones, and generates 15 authTags in total. Since on-chip storage is limited, feature maps and weights must be partitioned into smaller tiles, which creates overlapping regions between adjacent tiles. When these tiles are directly employed as authentication blocks for integrity verification, the overlapping regions are authenticated multiple times, incurring redundant cryptographic operations and repeated off-chip data accesses. Cross-layer dependencies further amplify this effect. Second, authentication metadata itself incurs substantial off-chip memory traffic. As shown in \Cref{figures:introduction_overview}(d), red and blue regions indicate the storage footprints of authentication tags and encrypted data, respectively; fm and ps denote feature maps and partial sums. Each authenticated data block is paired with a dedicated tag, which must be generated on-chip, stored off-chip, and later fetched back for integrity verification. For intermediate data iteratively updated across multiple computation stages, this store-and-reload procedure repeats each round, further exacerbating off-chip bandwidth pressure. These two inefficiencies are rooted in a mismatch between DNN-specific tiled computation patterns and the block-level security granularity assumed by general-purpose integrity mechanisms. These effects motivate a rethinking of scheduling exploration for secure DNN accelerators.

In this paper, we target these two overhead sources directly in existing secure DNN accelerators. Tile overlaps cause redundant authenticated encryption operations and excessive traffic for both data and authTags, while per-block integrity verification increases metadata traffic and strains already limited bandwidth. The key insight underlying our approach is that DNN inference follows a fully deterministic computational structure: the sequence, granularity, and reuse pattern of all tile accesses are known statically before execution begins. This determinism enables security operations to be co-optimized with computation scheduling at compile time, rather than handled obliviously at runtime. To exploit this, we propose {\design}, a scheduling framework for secure DNN accelerators that leverages the deterministic structure of inference workloads, as shown in \Cref{figures:introduction_overview}(e)(f).

The core novelty of {\design} resides in security-aware scheduling optimization, not underlying cryptographic engineering. It treats the existing authenticated encryption engine as a fixed hardware primitive, and optimizes the granularity of authentication operation invocations by exploiting the deterministic tiling structure of DNN inference. The main contributions of this paper are summarized as follows:

\begin{itemize}
    \item We propose a fusion-aware chunking strategy that exploits tiling-induced overlap structure to eliminate redundant security operations, thereby reducing both computational overhead and memory traffic.
    \item We introduce a hierarchical authentication mechanism that uses tile-based authentication for partial sums and chunk-based authentication for feature maps, thereby reducing off-chip metadata accesses.
    \item We develop {\design}, a scheduling framework that incorporates these optimizations while remaining compatible with existing DNN accelerator architectures.
    \item We evaluate {\design} on representative workloads and hardware configurations, showing an average performance improvement of $50.2\%$ and a $91.2\%$ reduction in memory traffic of security metadata over baselines.
\end{itemize}

The remainder of this paper is organized as follows. Section~\ref{section:background} reviews DNN accelerator execution and TEE-based security mechanisms. Section~\ref{section:motivation} presents the threat model and motivates the design by characterizing the overhead sources in existing secure DNN execution schemes. Section~\ref{section:framework} presents the {\design} framework in detail. Section~\ref{section:evaluation} evaluates {\design} against state-of-the-art baselines. Section~\ref{section:related_work} surveys related work, and Section~\ref{section:conclusion} concludes.
\section{Background} \label{section:background}

\subsection{Convolutional Neural Networks}

\begin{figure}
    \centering
    \includegraphics[width=0.4\textwidth]{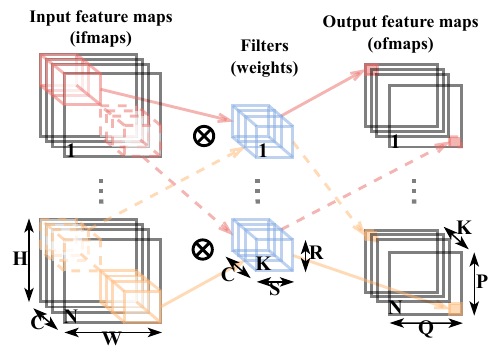}
    \caption{Basic CONV layer computation.}
    \label{figures:background_cnn}
    \vspace{-10pt}
\end{figure}

\begin{figure}
    \centering
\begin{lstlisting}[captionpos=b,caption={7D nested loop for CONV.}, label={figures:background_cnn_loop_list}, language=Python]
# P=(H-R+2*padding)/stride + 1
# Q=(W-S+2*padding)/stride + 1
for n in range(N): # N: number of inputs
  for k in range(K): # K: number of output channels
    for c in range(C): # C: number of input channels
      for p in range(P): # P: height of ofmaps
        for q in range(Q): # Q: width of ofmaps
          for r in range(R): # R: height of weights
            for s in range(S): # S: width of weights
              ofmaps[n][k][p][q] += 
                ifmaps[n][c][p+r][q+s] * 
                  weights[k][c][r][s]
\end{lstlisting}
    \vspace{-10pt}
\end{figure}

Convolutional Neural Networks (CNNs) are a popular architecture in DNNs, primarily consisting of multiple convolutional (CONV) layers that perform high-dimensional convolutions~\cite{krizhevsky2012imagenet,he2016deep,parashar2019timeloop}, as shown in \Cref{figures:background_cnn}. In many popular CNNs, CONV layers account for over 90\% of the overall operations and incur substantial data movement~\cite{krizhevsky2012imagenet,chen2016eyeriss}. The CONV, a core component of CNNs, efficiently extracts features from high-dimensional data through local connectivity and weight-sharing mechanisms. Specifically, a CONV layer employs learnable convolutional kernels that slide over the input data, performing localized dot product operations to generate feature maps that reflect the activation strength of specific patterns.

Mathematically, the convolutional operation can be expressed in a seven-dimensional nested-loop form~\cite{parashar2019timeloop,shrivastava2023securator,lee2023secureloop}, as shown in Listing~\ref{figures:background_cnn_loop_list}. In a CNN layer, the computational inputs consist of ifmaps and convolutional filter weight tensors. The output is generated as ofmaps through the convolution operation. The filter weights are represented as four-dimensional tensors, with dimensions organized as follows: spatial height (R), spatial width (S), number of input channels (C), and number of output channels (K). This organization follows the (R, S, C, K) convention commonly used in spatial convolution operations. Meanwhile, the ifmaps are structured as four-dimensional input tensors with dimensions specified as batch size (N), spatial height (H), spatial width (W), and number of input channels (C). The ofmaps are generated as four-dimensional output tensors with dimensions batch size (N), spatial height (P), spatial width (Q), and number of output channels (K), following the (N, P, Q, K) layout for output feature representation.

\subsection{TEEs for DNN accelerators}

\begin{figure}[!t]
    \centering
    \includegraphics[width=0.45\textwidth]{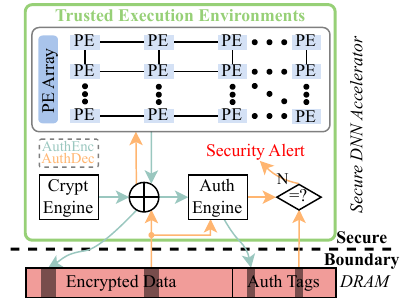}
    \caption{Diagram of TEEs for DNN accelerators.}
    \label{figures:background_tees}
    \vspace{-10pt}
\end{figure}

Trusted execution environments (TEEs) have emerged as a pivotal security paradigm for DNN accelerators, establishing hardware-enforced security boundaries between trusted on-chip components and untrusted off-chip peripherals~\cite{wang2022rt,yan2006improving, taassori2018vault,juvekar2018gazelle}. Contemporary TEE implementations employ cryptographic primitives, including memory encryption and integrity verification mechanisms, to ensure confidentiality and data integrity for information stored in external DRAM modules, as shown in \Cref{figures:background_tees}. Specifically, authenticated encryption schemes combining the Advanced Encryption Standard (AES) algorithms with Message Authentication Code (MAC) mechanisms provide real-time cryptographic protection against physical and software-based memory attacks.

\begin{figure}[!t]
    \centering
    \includegraphics[width=0.48\textwidth]{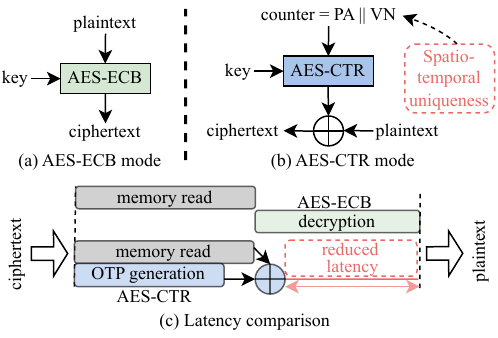}
    \caption{Comparison of AES encryption modes: (a) AES-ECB mode uses only a global encryption key. (b) AES-CTR mode uses a counter (PA $||$ VN) of spatio-temporal uniqueness to defend replay attacks. (c) AES-CTR mode reduces latency by overlapping memory access and OTP generation.}
    \label{figures:background_aes_diagram}
    \vspace{-10pt}
\end{figure}

A straightforward method for encrypting/decrypting data blocks is the electronic codebook mode (AES-ECB), as shown in \Cref{figures:background_aes_diagram}(a). However, AES-ECB mode is inherently unsuitable for cryptographic protocols because it encrypts blocks independently, which can lead to the exposure of data patterns~\cite{schneier2007applied}. This vulnerability allows attackers to infer sensitive plaintext (e.g., the presence of repeated data), which can be exploited in various attack scenarios. Worse still, memory reads lie on the critical path of program execution. In the AES-ECB decryption mode, due to its strictly serial execution nature, each memory read operation is followed by high latency decryption along the critical path, as shown in \Cref{figures:background_aes_diagram}(c). Another widely adopted implementation utilizes AES in counter mode (AES-CTR) for confidential data encryption, combined with Message Authentication Code (MAC) generation for integrity verification. The AES core engine generates a one-time pad (OTP) via counter-mode operations, which is then XORed with plaintext/ciphertext blocks for encryption/decryption processes, as shown in \Cref{figures:background_aes_diagram}(b). The counter concatenates the physical address (PA) of a data block that will be encrypted and the version number (VN), which is monotonically incremented on each write memory operation, to defend against replay attacks by integrating this spatio-temporal uniqueness information. In the encryption path, this OTP undergoes a bitwise XOR operation with plaintext blocks to produce ciphertext outputs. Conversely, during decryption, the identical OTP (regenerated through seed replication) is XORed with the received ciphertext to reconstruct the original plaintext through deterministic reversal. By overlapping memory access and OTP generation, AES-CTR mode significantly reduces latency, as shown in \Cref{figures:background_aes_diagram}(c). This pipelined execution allows for improved throughput, which is particularly beneficial in DNN accelerators where high throughput is essential for efficient processing of large datasets.

Existing approaches fall into two categories: coarse-grained tile-based authentication and fine-grained slice-based authentication for integrity verification. Tile-based methods treat each tile as an independent authentication block, leading to redundant security operations in overlapping regions. Slice-based methods attempt to mitigate overlaps by dividing data into smaller slices, but significantly increase exploration-space complexity and may introduce additional authentication metadata overhead. Consequently, the tight coupling between security block granularity and DNN computation patterns creates a challenging optimization problem, forcing existing methods to rely on heuristic searches that often yield suboptimal solutions. To address these challenges, it is essential to rethink scheduling exploration for efficient secure DNN acceleration.

\section{Threat Model and Motivation} \label{section:motivation}

\subsection{Threat Model} \label{section:motivation:threat_model}

The threat model for our work encompasses potential attacks on the integrity and confidentiality of data processed by deep neural networks (DNNs). Similar to the threat model proposed in prior literature~\cite{gassend2003caches,hua2022mgx, lee2022tnpu,xuan2025seda, 2026arXiv260220521X}, we assume that the accelerator itself operates as a trusted computing base, whose internal state cannot be directly observed or tampered with. Conversely, any external device connected to the accelerator is considered untrusted, such as off-chip memory and communication buses, meaning that attackers could potentially access or manipulate any information stored in these devices through physical or software attacks. Our scope excludes other physical side-channel attacks for inferring DNN architectures~\cite{yan2020cache}, such as power and electromagnetic side channels. Adversarial attacks against machine learning algorithms are also outside the scope of this work~\cite{akhtar2018threat}, as our focus is on protecting memory confidentiality and integrity during accelerator execution.

\subsection{Redundant Security Operations from Tile Overlaps} \label{section:motivation:tile_overlaps}

\begin{figure}[!t]
    \centering
    \includegraphics[width=0.4\textwidth]{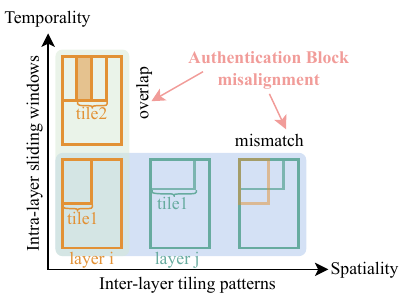}
    \caption{Misalignment between data tiles and authentication blocks, which arises from intra-layer sliding windows and inter-layer tiling patterns.}
    \label{figures:framework_spatial_temporal}
    \vspace{-10pt}
\end{figure}

Tile overlaps in DNN accelerators pose significant challenges for secure scheduling, leading to redundant security operations and increased memory traffic. As illustrated in \Cref{figures:framework_spatial_temporal}, secure DNN accelerators face critical challenges from both intra-layer sliding window overlaps and inter-layer tiling discrepancies, resulting in misalignment between authentication blocks and data tiles. Intra-layer sliding window overlaps, shown along the temporal axis, occur during convolution or pooling operations when the kernel stride is smaller than its spatial dimension, causing consecutive computation windows to share partial data and requiring redundant authentication operations. Inter-layer tiling-induced overlaps, depicted along the spatial axis, emerge from mismatched data partitioning strategies across consecutive layers, where different tiling patterns (tile size, stride, or padding) prevent the reuse of authentication metadata. For instance, when one layer processes data in large tiles while the subsequent layer uses smaller tiles, the authentication metadata generated for larger tiles cannot be directly reused, necessitating additional computations and data transfers. Such tiling misalignments underutilize on-chip scratchpad buffer and exacerbate off-chip memory traffic, motivating the development of fusion-aware scheduling strategies that align authentication granularity with computational tiling patterns.

\subsection{Authentication Metadata Access Overhead}

In secure DNN accelerators, each data block requires authTags (e.g., MACs or hash values) to ensure data integrity and prevent tampering attacks. These authentication metadata must be stored and transmitted alongside data blocks, thereby introducing significant memory access overhead. While authentication metadata for read-only inputs and weights can be pre-allocated according to dataflow patterns, frequently updated data present considerably more complex challenges. In particular, partial sums and feature maps require dynamic authTag generation after each computation, which leads to substantial overhead in memory-bound systems due to frequent off-chip DRAM accesses. This overhead is further exacerbated by the high volume of data processed in DNNs and the computational cost of generating and storing authentication metadata for each operation. To effectively mitigate these issues, efficient scheduling strategies must minimize authentication metadata accesses while maintaining data integrity. This can be achieved by leveraging the deterministic nature of DNN computations to enable optimized dataflow patterns that reduce frequent authTag generation. Moreover, hierarchical layer authentication mechanisms can manage metadata more effectively, ensuring that only necessary accesses occur while simultaneously minimizing memory traffic and computational overhead.

\section{{\design} Framework} \label{section:framework}

\subsection{Overview of Proposed {\design}} \label{section:framework:overview}

\begin{figure}[!t]
    \centering
    \includegraphics[width=0.48\textwidth]{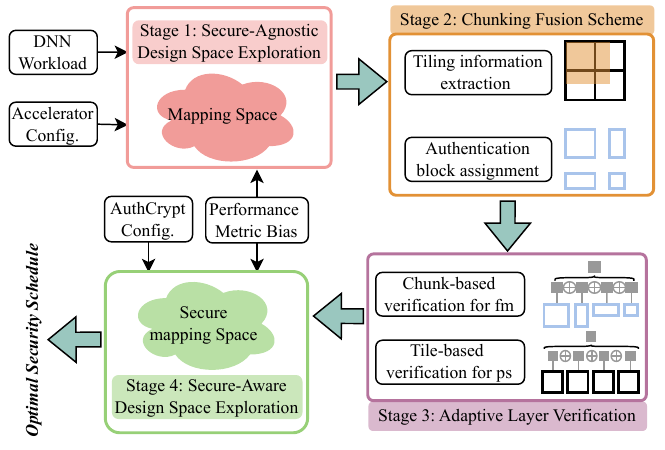}
    \caption{Overview of {\design} framework.}
    \label{figures:framework_sesa}
    \vspace{-10pt}
\end{figure}

In this section, we present {\design}, an efficient and practical scheduling framework for secure DNN accelerators. The core premise of {\design} is that DNN inference follows a statically deterministic computation structure, which makes it possible to analyze all tile access patterns, overlap boundaries, and intermediate data flows before execution. This static predictability enables security operations to be co-optimized with computation scheduling at compile time rather than handled obliviously at runtime.

As shown in \Cref{figures:framework_sesa}, the framework consists of four stages that progressively refine a secure schedule. \textit{Stage 1: Secure-Agnostic Design Space Exploration} explores mappings optimized for performance without security constraints, producing a candidate design space that serves as the starting point for subsequent security-aware optimization. This stage is orthogonal to existing DSE methodologies~\cite{parashar2019timeloop, accelergy}: we use standard DSE tools to generate candidate mappings for a given workload and accelerator, decoupling performance exploration from security policy. \textit{Stage 2: Fusion-Aware Chunking Scheme} analyzes the tiling structures of each candidate mapping and identifies inter-tile overlap regions across layers. Authentication block boundaries are then assigned to align with these overlap regions, grouping overlapped data into fusion-aware chunks to eliminate redundant authentication operations. \textit{Stage 3: Adaptive Layer-Based Verification} builds on the chunk partitioning from Stage~2 and applies a differentiated authentication policy: chunk-based verification for feature maps---where cross-layer tiling misalignment prevents direct block reuse---and tile-based verification for partial sums---where read and write boundaries remain aligned within a single layer. This hierarchical strategy reduces off-chip metadata traffic while preserving strong integrity guarantees. \textit{Stage 4: Security-Aware Design Space Exploration} re-evaluates the candidate mappings under their actual security costs incurred after Stages~2 and~3, selecting the final mapping that optimally balances security overhead and runtime performance. Because all four optimizations are introduced at the scheduling level, the framework preserves the conventional tile-level execution interface of the underlying accelerator.

The framework operates iteratively, allowing information from later stages to refine earlier design choices until the schedule converges. We begin with the chunking stage, since overlap-induced redundancy is the first source of security overhead that the framework addresses. Compared to prior scheduling-aware secure DNN work such as SecureLoop~\cite{lee2023secureloop}, which treats overlap as a constraint to be managed via heuristic search, {\design}'s key distinction is its elimination of overlap-induced redundancy through compile-time analysis.

\subsection{Fusion-Aware Chunking Scheme}

\subsubsection{Misalignment Characteristics} \label{section:framework:fusion:misalignment}

\begin{figure}[!t]
    \centering
    \includegraphics[width=0.5\textwidth]{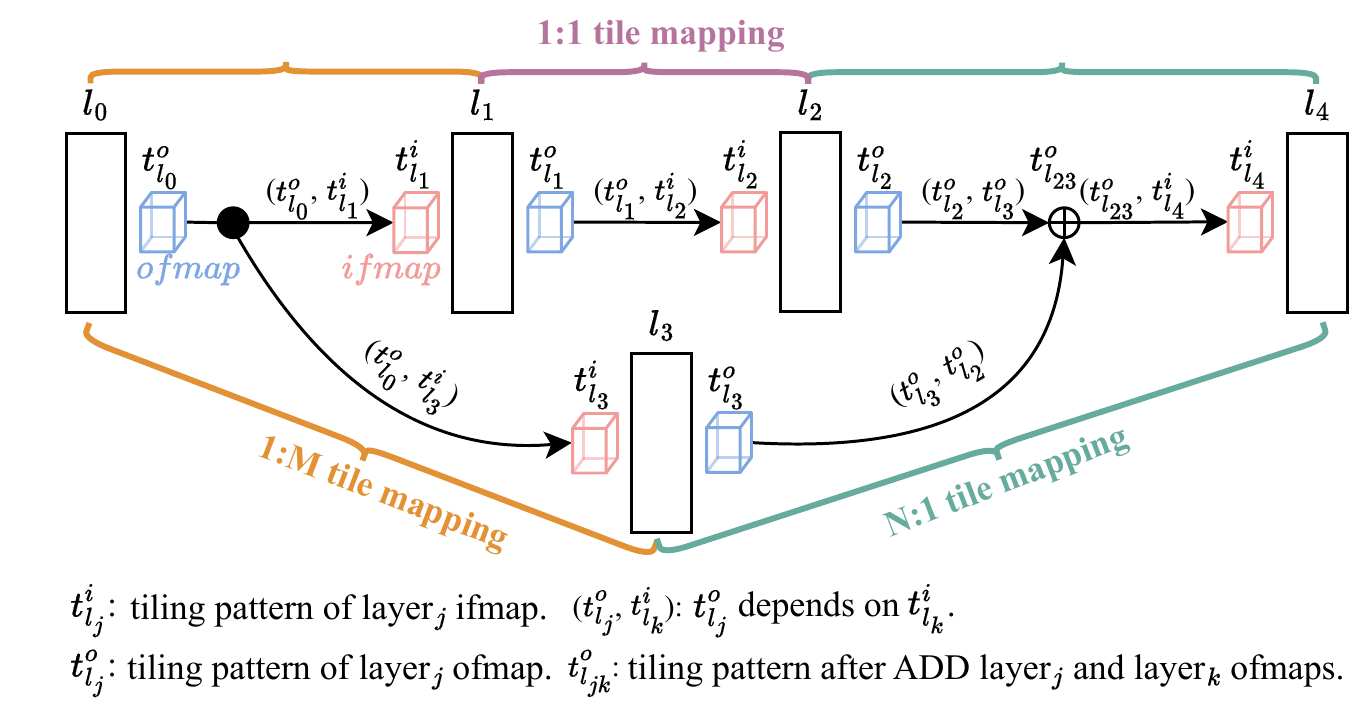}
    \caption{Three types of inter-layer tiling mapping dependencies.}
    \label{figures:framework_tile_mapping_types}
    \vspace{-10pt}
\end{figure}

Misalignment between the tiling structures of adjacent layers leads to inefficient data access and unnecessary security overhead. As shown in \Cref{figures:framework_spatial_temporal}, this misalignment arises from two sources. First, intra-layer sliding-window movement creates overlaps between tiles within the same layer; if each tile is authenticated independently, overlapped regions are processed repeatedly. Second, inter-layer tiling discrepancies create cross-layer mismatches that prevent direct reuse of authentication metadata. We address both effects using a fusion-aware chunking scheme that aligns authentication block boundaries with tile boundaries, thereby minimizing redundant computations and memory accesses.

Intra-layer sliding window techniques dynamically adjust tile boundaries within a single layer to optimize data access patterns, as shown in \Cref{figures:framework_spatial_temporal} along the temporal dimension. Although such approaches enhance memory utilization and reduce latency, the resulting tile overlaps introduce redundant security operations and increase memory traffic overhead.

Inter-layer tiling relationships are categorized into 1:1, 1:M, and N:1 mappings, as shown in \Cref{figures:framework_tile_mapping_types}: (1) 1:1 tile mapping, in which one tiling pattern of layer$_i$ ofmap depends on one tiling pattern of layer$_j$ ifmap, such as $t^o_{l_1}$ and $t^i_{l_2}$; (2) 1:M tile mapping, in which one tiling pattern may associate with multiple downstream tiling patterns, such as $t^o_{l_0}$ with $t^i_{l_1}$ and $t^i_{l_3}$; and (3) N:1 tile mapping, in which multiple upstream tiling patterns are associated with one downstream tiling pattern, such as ($t^o_{l_2}$, $t^o_{l_3}$) and $t^i_{l_4}$. Both 1:M and N:1 inter-layer mapping cases can be further decomposed into multiple 1:1 mappings to streamline the optimization process, which introduce data fragmentation when distinct tiling strategies coexist. A precomputed chunk boundary, derived at compile time from the union of all overlapping boundaries, maps each fragmented region to a unique logical chunk identifier, ensuring each physical block is authenticated exactly once regardless of the mapping type.

\subsubsection{Model of Chunking as Authentication Block} Chunking transforms tile structures into authentication blocks corresponding to overlapping or non-overlapping regions, eliminating redundant security operations and memory traffic. Our chunk-based authenticated cryptographic engine (AuthCrypt) performs authentication and encryption separately for overlapping and non-overlapping regions while preserving tile-level data blocks for DRAM communication and PE Array computation.

\begin{figure}[!t]
    \centering
    \includegraphics[width=0.48\textwidth]{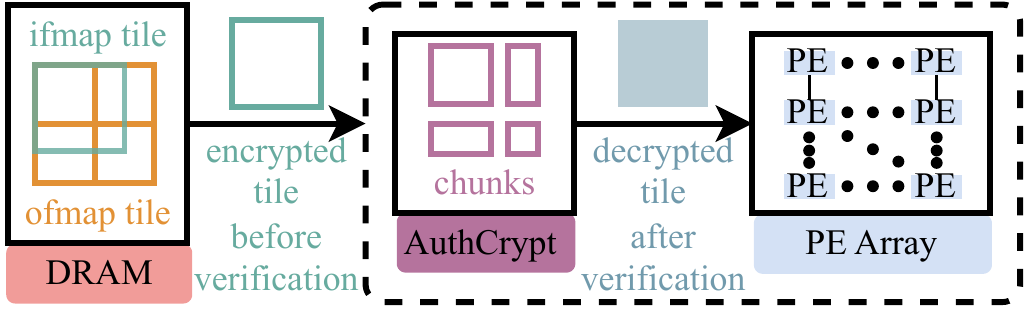}
    \caption{Chunk-based authentication block organization for authenticated encryption in secure DNN accelerators.}
    \label{figures:framework_chunk}
    \vspace{-10pt}
\end{figure}

\Cref{figures:framework_chunk} illustrates how chunk-based authentication is applied during authenticated decryption, when a DNN accelerator fetches feature maps from untrusted off-chip DRAM for inference computation. The left side of the figure depicts the raw tile layout, where the ofmap tiles of layer$_i$ and the ifmap tiles of layer$_j$ do not share the same boundaries. If these tiles are used directly as authentication blocks for integrity verification, overlapping regions are authenticated multiple times, resulting in redundant cryptographic computations and unnecessary off-chip traffic. The right side of the figure depicts the internal operation of our AuthCrypt module. Upon receiving a tile access request, AuthCrypt identifies chunk-boundaries to decompose the tile into its constituent overlapping and non-overlapping chunks. Each chunk is then authenticated independently only once and their tags are stored on-chip for reuse when the same data is accessed by the adjacent layer. Externally, AuthCrypt preserves the conventional tile-level interface for both DRAM transfers and PE-array execution, ensuring backward compatibility with existing accelerator pipelines. Internally, it operates at chunk granularity to eliminate redundant security work. Because the tiling behavior of a mapped DNN is fully determined at compile time, all chunk boundaries can be precomputed from the mapping.

\begin{figure}[!t]
    \centering
    \includegraphics[width=0.48\textwidth]{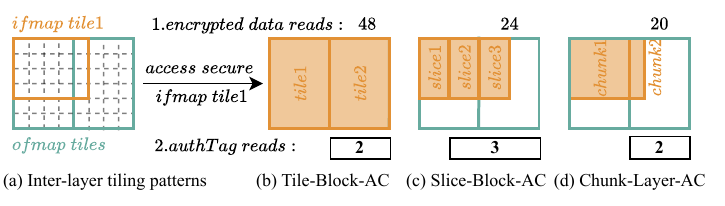}
    \caption{Transforming inter-layer tiling discrepancies (a) with different scheduling schemes (b-d).}
    \label{figures:framework_comp_chunk}
    \vspace{-10pt}
\end{figure}

Modeling chunking as authentication blocks reduces security overhead without changing the tile-level communication interface. In our framework, \textit{chunking converts the original tile structure into contiguous authentication blocks derived from spatial and temporal overlap relationships}. The chunk boundaries are determined by the access patterns of the mapped workload, allowing authentication granularity to adapt to cross-layer interactions while preserving tile-level transfers for computation and DRAM communication.
To make the effect of this transformation concrete, we next compare the three scheduling schemes under the same inter-layer mismatch scenario.

\subsubsection{Example of Comparison}

\Cref{figures:framework_comp_chunk}(a) illustrates the root cause of the overhead: the ofmap tiles of layer$_i$ and the ifmap tiles of layer$_j$ use different tiling patterns, so producing and consuming views of the same data do not share aligned boundaries. Any scheme that treats tiles directly as authentication blocks must therefore handle the same physical data multiple times under different tile identities, incurring both redundant encrypted-data reads and repeated authTag accesses. \Cref{figures:framework_comp_chunk}(b--d) compares three authentication scheduling schemes under this tiling misalignment scenario: the baseline Tile-Block-AC, state-of-the-art Slice-Block-AC, and our proposed Chunk-Layer-AC (detailed in \Cref{section:exp:scheduling_framework}).

As shown in \Cref{figures:framework_comp_chunk}(b), Tile-Block-AC uses tile-sized authentication blocks that are directly inherited from the computation tiling. For the target ifmap tile1 access, this requires 48 encrypted data reads and 2 authTag reads, because the tile boundaries of layer$_i$'s ofmaps do not align with the accessed region of layer$_j$'s ifmaps, forcing multiple full-tile fetches to cover the required data. As shown in \Cref{figures:framework_comp_chunk}(c), Slice-Block-AC decomposes tiles into finer horizontal slices to reduce encrypted data reads to 24, halving the data traffic. However, finer slicing increases the number of distinct authentication blocks, raising authTag overhead to 3 reads. Moreover, slice boundaries are still not aligned with the true overlap structure, so some redundancy remains.

\begin{figure}[!t]
    \centering
    \includegraphics[width=0.4\textwidth]{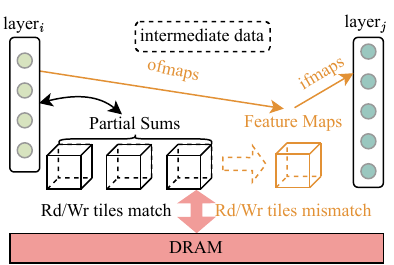}
    \caption{Tile boundary match analysis for read and write operations on feature maps (fm) and partial sums (ps).}
    \label{figures:framework_fm_ps}
    \vspace{-10pt}
\end{figure}

As shown in \Cref{figures:framework_comp_chunk}(d), our proposed Chunk-Layer-AC derives authentication block boundaries directly from the inter-layer overlap structure. Overlapping and non-overlapping regions are identified at compile time and mapped to separate chunks whose boundaries precisely match the cross-layer access pattern. As a result, each region is read and authenticated exactly once, reducing encrypted data reads to 20---the lowest among all three schemes---while keeping authTag reads at 2, matching Tile-Block-AC without its data redundancy. This example highlights the central advantage of chunk-aligned authentication: by removing boundary-induced redundancy at the source, it simultaneously \textit{reduces both data traffic and security metadata traffic without weakening integrity protection}. Because the chunk boundaries are derived statically from the mapped workload, the same optimization applies uniformly to all deterministic layer interactions in the network.

\subsection{Adaptive Layer-Based Verification}

\subsubsection{Feature Maps and Partial Sums}

In DNN accelerators, intermediate data can be categorized by their cross-layer behavior: \textit{partial sums are intra-layer intermediate results computed within a single layer, whereas feature maps are inter-layer outputs transferred between consecutive layers}. Both require integrity verification and are typically tile-organized, but their different boundary-alignment properties lead to different sources of security overhead. We therefore employ a hierarchical authentication strategy: tile-based verification for partial sums and chunk-based verification for feature maps.

\Cref{figures:framework_fm_ps} illustrates the tile-boundary analysis for read and write operations on intermediate data in DNN accelerators, highlighting the different authentication requirements of partial sums and feature maps. The figure depicts two consecutive layers (layer$_i$ and layer$_j$) that exchange intermediate data stored in DRAM. For partial sums (shown in black on the left), the read and write boundaries remain aligned with layer$_i$. The three cubic tiles indicate that both operations use consistent tile boundaries, enabling tile-based authentication without extra remapping overhead because the same authentication blocks can be reused. In contrast, for feature maps (shown in orange on the right), the read and write boundaries are misaligned across consecutive layers. The orange arrows indicate that the ofmaps from layer$_i$ become the ifmaps for layer$_j$, but the tiling patterns differ due to varying computational requirements, kernel sizes, or parallelization strategies. This mismatch prevents direct reuse of tile-level authentication metadata, because the data written by one layer do not match the data accessed by the next. This observation motivates the hierarchical authentication strategy used in our framework: \textit{tile-based verification for partial sums and chunk-based verification for feature maps}.

\subsubsection{Adaptive Layer Verification Scheme}

\begin{algorithm}[t]
    \DontPrintSemicolon
    \LinesNumbered
    \SetKwInOut{Input}{\textbf{Input}}
    \SetKwInOut{Output}{\textbf{Output}}
    \SetKwFunction{algo}{algo}
    \SetKwFunction{proc}{Procedure}{}{}
    \SetKwFunction{Master}{\textbf{Master}}{}{}
    \SetKwFunction{main}{\textbf{ChooseCenter}}
    \Input{(1) $feature\;maps$ or $partial \; sums$: encrypted intermediate data; (2) PA, VN: physical address, version number.\\}
    \Output{$auth\_layer\_block$: layer authentication.\\}

    \If{ feature maps}{ \label{alg:layer_veri_line_if_start}
        \For{ tile in feature maps}{
            \For{ chunk in tile}{
                $auth\_chunk\_block \leftarrowtail \textbf{\textsc{CalcAuth}}(enc\_chunk\_block, chunk\_id);$\label{alg:layer_veri_line_auth_chunk_block}\\
                $auth\_tile\_block \leftarrowtail auth\_tile\_block \oplus auth\_chunk\_block;$ \label{alg:layer_veri_line_auth_tile_block} \\
            }
            $auth\_layer\_block \leftarrowtail auth\_layer\_block \oplus auth\_tile\_block;$ \label{alg:layer_veri_line_auth_layer_block} \\
        }
    }
    \Else{ \label{alg:layer_veri_line_else_start}
        \For{ tile in partial sums}{
            $auth\_tile\_block \leftarrowtail \textbf{\textsc{CalcAuth}}(enc\_tile\_block, tile\_id);$ \label{alg:layer_veri_line_auth_tile_block2}  \\
            $auth\_layer\_block \leftarrowtail auth\_layer\_block \oplus auth\_tile\_block;$ \label{alg:layer_veri_line_auth_layer_block2} \\
        }
    }

    \SetKwBlock{Function}{\texttt{function} \textsc{CalcAuth}($enc\_data\_block, data\_id$)}{end}  \label{alg:layer_veri_line_CalcAuth}

    \Function{
        $auth\_tag \leftarrowtail Hash(\text{PA} || data\_id ||\text{VN}|| enc\_data\_block);$ \label{alg:layer_veri_line_auth_tag} \\
    }

    \caption{Layer-based verification scheme.} \label{alg:layer_verification}
\end{algorithm}

Algorithm~\ref{alg:layer_verification} formalizes the layer-based hierarchical verification scheme. The algorithm takes as input encrypted intermediate data (feature maps or partial sums) along with their physical addresses (PA) and version numbers (VN), and produces a single layer-level authentication block that compactly represents the integrity state of the entire layer. Deferring verification to layer granularity is the key mechanism for reducing off-chip metadata traffic: rather than storing and retrieving one authTag per data block during every computation round, the scheme accumulates chunk- or tile-level tags on-chip via XOR aggregation and only commits a single layer-level tag to DRAM.

The algorithm applies a differentiated policy based on data type. For feature maps (lines~\ref{alg:layer_veri_line_if_start}--\ref{alg:layer_veri_line_auth_layer_block}), authentication proceeds at chunk granularity to match the cross-layer overlap structure identified by the chunking scheme. For each tile, the inner loop iterates over its constituent chunks, invoking \textsc{CalcAuth} with the encrypted chunk block and its $chunk\_id$ to produce a chunk-level tag (line~\ref{alg:layer_veri_line_auth_chunk_block}). These chunk tags are XOR-aggregated into a tile-level block $auth\_tile\_block$ (line~\ref{alg:layer_veri_line_auth_tile_block}), which is in turn XOR-folded into the running layer-level block $auth\_layer\_block$ (line~\ref{alg:layer_veri_line_auth_layer_block}). For partial sums (lines~\ref{alg:layer_veri_line_else_start}--\ref{alg:layer_veri_line_auth_layer_block2}), whose read and write boundaries remain aligned within a single layer, authentication is performed directly at tile granularity without the intermediate chunk loop. \textsc{CalcAuth} is invoked once per tile using the $tile\_id$ (line~\ref{alg:layer_veri_line_auth_tile_block2}), and the resulting tile-level tags are XOR-aggregated into the layer-level tag (line~\ref{alg:layer_veri_line_auth_layer_block2}).

The shared \textsc{CalcAuth} subroutine (lines~\ref{alg:layer_veri_line_CalcAuth}--\ref{alg:layer_veri_line_auth_tag}) generates an authTag by hashing the concatenation of the physical address (PA), data identifier ($data\_id$), version number (VN), and encrypted data block. PA provides spatial uniqueness to bind each tag to a specific memory location, while VN provides temporal uniqueness to prevent replay attacks in which a stale but valid ciphertext is substituted for a current value. Together, this design matches authentication granularity to the actual data access pattern of each data type, reducing off-chip metadata traffic without compromising the integrity guarantees required by the threat model.

\begin{figure}[!t]
    \centering
    \includegraphics[width=0.48\textwidth]{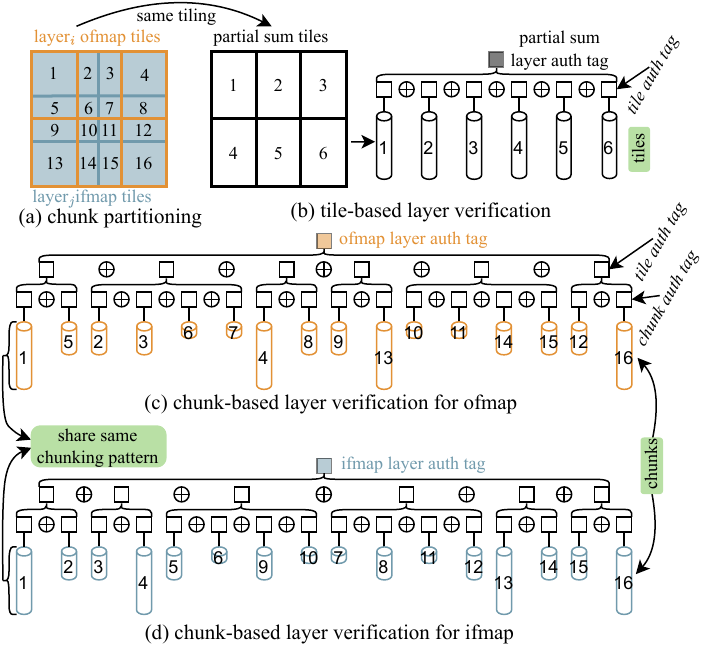}
    \caption{Adaptive layer-based verification case.}
    \label{figures:framework_layer_verify}
    \vspace{-10pt}
\end{figure}

\subsubsection{Example of Layer-Based Verification}

\Cref{figures:framework_layer_verify} presents a concrete example of the adaptive layer-based verification scheme, illustrating how feature maps and partial sums are authenticated under the hierarchical strategy.

\Cref{figures:framework_layer_verify}(a) shows the chunk partitioning for feature maps. Because the ofmap tiling pattern of layer$_i$ differs from the ifmap tiling pattern of layer$_j$, the two views of the shared data are fused to derive a unified chunk layout. The shared data region is partitioned into 16 chunks whose boundaries align with the union of both tiling patterns, ensuring that each chunk is a contiguous, non-overlapping unit visible to both layers. Each chunk is processed exactly once to generate an authTag, eliminating redundant authentication across the layer boundary.

\Cref{figures:framework_layer_verify}(b) shows the tile-based authentication for partial sums. Since partial sums are intra-layer intermediate results, their read and write operations both use the tiling pattern of layer$_i$, keeping the tile boundaries consistent. No chunking is required; each tile is authenticated directly using \textsc{CalcAuth}, and all tile-level tags are XOR-aggregated to obtain the layer-level authTag. Each time the complete set of partial sums for a layer is written back to DRAM, the version number VN is incremented to ensure temporal freshness. Because the data written and subsequently read are identical, the layer authTags computed on the write path and the read path remain consistent, confirming the integrity of the partial-sum data without additional metadata overhead.

\Cref{figures:framework_layer_verify}(c) and \Cref{figures:framework_layer_verify}(d) show how the same chunk layout supports integrity verification from both the producing and consuming perspectives. In \Cref{figures:framework_layer_verify}(c), layer$_i$ writes its ofmap tiles to DRAM; the AuthCrypt module decomposes each ofmap tile into its constituent chunks according to the fused partition, computes a chunk-level tag for each, and XOR-aggregates them into a layer authTag $T_{\mathrm{write}}$. In \Cref{figures:framework_layer_verify}(d), layer$_j$ reads the same data as ifmap tiles; although the tile-to-chunk decomposition differs from layer$_i$'s perspective, both views operate on the same underlying chunk data blocks. Each chunk carries a unique spatial identifier ($\mathrm{PA} \| chunk\_id$) and a temporal marker (VN), so the \textsc{CalcAuth} outputs are identical regardless of the access order. The XOR-aggregated result yields a layer authTag $T_{\mathrm{read}}$. When $T_{\mathrm{read}} = T_{\mathrm{write}}$, the integrity of the feature map data across the layer boundary is confirmed. This cross-layer consistency property is a direct consequence of the XOR aggregation's commutativity: reordering the chunks does not change the aggregate tag, so both layers converge to the same verification result without requiring synchronized access order.

\subsection{Security Analysis}

{\design} is a scheduling-level optimization. It does not replace the underlying cryptographic primitives but restructures how and when they are applied. We show that all security properties required by the threat model in \Cref{section:motivation:threat_model} are fully preserved.

\subsubsection{Data Confidentiality}
Every off-chip data block is encrypted with AES-CTR. The counter is seeded by $(\text{PA} \| \text{VN})$, binding each ciphertext block to a unique memory location and write epoch~\cite{gassend2003caches,yan2006improving}. {\design} changes only the granularity at which blocks are defined. Each chunk inherits its own counter value derived from its physical address and the current VN. AES-CTR is a stream cipher whose security holds independently for each block. Reducing block size to chunk granularity therefore strictly preserves the semantic security of the encryption. Chunk boundaries are determined entirely by static tiling analysis at compile time. They carry no secret information, so their exposure to an adversary does not assist decryption.

\subsubsection{Security Assumptions of XOR Aggregation} \label{section:framework:verification:xor}

We provide a formal security argument for the layer-level aggregate $T_{\text{layer}} = \bigoplus_i t_i$, where each tag is computed as $t_i = \mathcal{H}(K, \text{PA}_i, data\_id_i, \text{VN}_i, enc\_data\_block_i)$ and $\mathcal{H}$ is a collision-resistant keyed hash. Since each $t_i$ is uniquely bound to its block, any modification to a block's ciphertext, address, or version number changes $t_i$ to $t_i' \neq t_i$ with probability at least $1 - 2^{-m}$ (for an $m$-bit tag). The resulting aggregate becomes $T_{\text{layer}}' = T_{\text{layer}} \oplus t_i \oplus t_i'$, which differs from $T_{\text{layer}}$ unless $t_i \oplus t_i' = 0$, an event that occurs only with negligible probability. For multi‑block tampering, the adversary would need to find a set $S$ such that $\bigoplus_{i \in S} t_i' = \bigoplus_{i \in S} t_i$ without knowing $K$. This is provably as hard as breaking the MAC itself under standard assumptions. Therefore, the XOR aggregation preserves the integrity guarantees of the individual tags and does not introduce any new security vulnerabilities. This aggregation technique is consistent with prior work on compact integrity verification~\cite{shrivastava2023securator,xuan2025seda} and has been proven secure in incremental cryptographic settings.

\subsubsection{Replay Attack Resistance}
Replay attacks substitute a stale but previously valid ciphertext block for a current one. {\design} prevents this through the version number VN, which is monotonically incremented on each write operation. A replayed block carries an outdated VN. As a result, the regenerated OTP differs from the one used at write time, and decryption produces incorrect plaintext. Simultaneously, the stale VN causes \textsc{CalcAuth} to produce a tag that disagrees with the current layer-level aggregate. This mismatch is detected independently at verification time~\cite{lee2022tnpu,hua2022mgx}. VNs are maintained entirely on-chip within the trusted domain. They therefore cannot be manipulated by an adversary whose access is limited to off-chip memory or the communication bus.

\subsubsection{Security Guarantees of Deferred Verification} \label{section:evaluation:framework:security:deferred}

We further strengthen the security analysis by formalizing the guarantees of our deferred verification mechanism from two perspectives. First, all unauthenticated plaintext, including intermediate states, remains strictly within the on-chip trusted domain. This requirement does not prohibit intermediate results from being written back to off-chip DRAM. Intermediate data may cross the trusted boundary as long as they are encrypted and integrity-protected before being transferred to untrusted memory. Any off-chip tampering is detected by a layer-level authTag mismatch, causing the affected outputs to be discarded before propagation. Second, regarding resilience against chosen-ciphertext attacks, our design exposes no decryption oracle to adversaries with only off-chip access, as all decrypted values and on-chip states remain unobservable. For resilience against chosen-ciphertext attacks (CCA), we note that a practical CCA attack requires access to a decryption oracle that reveals the plaintext corresponding to adversary-chosen ciphertexts. In our design, decrypted plaintext and on-chip intermediate states remain fully unobservable to adversaries with only off-chip memory and bus access privileges, such that no exploitable decryption oracle is exposed to the adversary.

\subsubsection{Execution Data Flow} \label{section:evaluation:framework:security:flow}

We clarify that deferred verification refers to deferring the final layer-level integrity check and off-chip metadata persistence to layer completion, rather than postponing per-block authentication or decryption. On the memory read path, the AuthCrypt module performs streaming decryption upon data arrival, generating decryption keystreams from physical addresses and version numbers in parallel with memory access. Decrypted plaintext is directly forwarded to the PE array for computation without waiting for authentication, while the authentication datapath concurrently computes per-block tags and accumulates them via commutative XOR into a layer-level authTag. After all tiles of a layer have been processed, the aggregated on-chip authTag is compared against the single layer-level authTag retrieved from DRAM; a match permits writeback of computation results with a newly generated layer-level authTag, while a mismatch aborts execution and invalidates all intermediate results to prevent tampered data from propagating. The write path follows a symmetric flow, where per-block tags are aggregated on-chip and only the final layer-level authTag is stored off-chip, eliminating per-block metadata traffic.

\subsection{Generalizability} \label{section:framework:general}


Although our evaluation uses representative CNN workloads, {\design}'s core mechanisms—fusion-aware chunking and hierarchical authentication—are applicable beyond CNNs. They rely solely on the observation that DNN inference follows a fully deterministic tiled execution structure, a property shared by nearly all DNN accelerator mappings regardless of model type. Transformers employ tiling for attention and matrix multiplication; and graph neural network (GNN) accelerators leverage vertex-tiling and sparse tiling formats. {\design}'s compile-time chunk-boundary analysis applies to any workload where tile access patterns are statically known. The primary limitation is for models with runtime-variable tiling (e.g., variable sequence lengths in transformers, or dynamic graph topologies in GNNs), which would require runtime reconfiguration—an interesting direction for future work.

{\design}'s chunking scheme is agnostic to layer types, because its optimization is driven entirely by data tiling patterns and cross-layer access relationships, rather than by specific operator semantics. For element-wise operations (e.g., non-ReLU activations), tiling boundaries remain unchanged; {\design} therefore fuses them with the preceding computation and does not introduce new authentication boundaries. For sliding-window operations (e.g., pooling layers), the window overlaps are structurally isomorphic to those of convolutions and are directly integrated into the same cross-layer chunking framework. For residual connections (representative N:1 tiling mappings), we derive unified chunk boundaries by taking the union of tiling patterns from all input branches and the output, ensuring that the converged data is authenticated only once.

\subsection{Conditions with Limited Benefit} \label{section:framework:benefit}

{\design}'s performance gains, achieved by eliminating overlap-induced redundant authentication and cutting off-chip authentication metadata traffic, scale with tile overlap severity and memory bandwidth pressure. Fusion-aware chunking yields diminishing returns for operators without sliding-window overlap (e.g., fully-connected layers, stride-matching convolutions) or perfectly aligned cross-layer tiling, yet incurs near-zero runtime overhead as all boundaries are precomputed at compile time. Hierarchical verification delivers smaller latency improvements when on-chip SRAM fits full-layer feature maps and tags, or when the accelerator is compute-bound—bottlenecked by processing elements and cryptographic throughput rather than DRAM bandwidth—though it still reduces total cryptographic operations for consistent energy savings. Scenarios with negligible benefits from both optimizations are rare in practical memory-bound inference deployments. {\design} can assess optimization gains and can fall back to baseline tile-level authentication to avoid any performance penalty.

\section{Evaluation} \label{section:evaluation}

\begin{table}[t]
    \caption{Summary of security frameworks. AC=Authenticated Cryptography; \xmark=no, \cmark=yes; \CIRCLE=high, \LEFTcircle=moderate, \Circle=no.}
    \label{table:comparison}
    \centering
    \footnotesize
    \setlength{\tabcolsep}{2pt}
    \begin{tabular}{c|c|c|c}
        \toprule
        \thead{Scheduling                               \\ framework} & \thead{AuthCrypt \\ granularity}  & \thead{Layer-based \\ integrity verification} & \thead{Redundant security\\ computation and traffic}\\
        \midrule
        Tile-Block-AC  & tile  & {\xmark} & \CIRCLE     \\
        Slice-Block-AC & slice & {\xmark} & \LEFTcircle \\
        Chunk-Layer-AC & chunk & {\cmark} & \Circle     \\
        \rowcolor[HTML]{EFEFEF}
        \bottomrule
    \end{tabular}
    \vspace{-10pt}
\end{table}

\subsection{Experimental Setup}

\subsubsection{Scheduling Framework} \label{section:exp:scheduling_framework}

We compare {\design} (denoted as Chunk-Layer-AC) with two representative secure scheduling frameworks: Tile-Block-AC (e.g., \cite{hua2022mgx, shrivastava2023securator, lee2022tnpu}) and Slice-Block-AC (e.g., \cite{lee2023secureloop}). The key differences between these frameworks are summarized in Table~\ref{table:comparison}. Tile-Block-AC authenticates at tile granularity, whereas Slice-Block-AC uses finer-grained slices to reduce overlap-induced redundancy. {\design} instead combines chunk-based authentication with hierarchical verification: chunk-based authentication for feature maps removes redundant security work caused by overlap, while tile-based authentication for partial sums avoids unnecessary authentication-metadata overhead.

\begin{figure*}[!t]
    \centering
    \subfloat[Normalized latency overhead. \label{figures:evaluation_overall_latency}]{\includegraphics[width=0.32\linewidth]{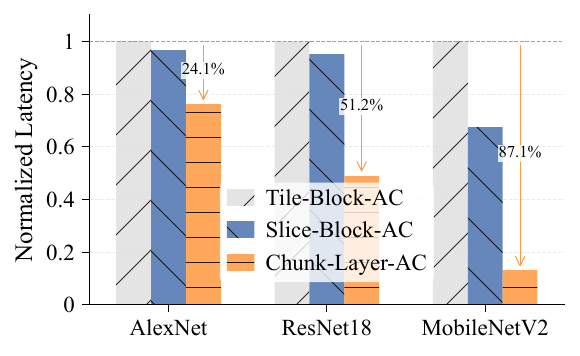}}
    \hfill
    \subfloat[Normalized additional memory traffic. \label{figures:evaluation_overall_traffic}]{\includegraphics[width=0.32\linewidth]{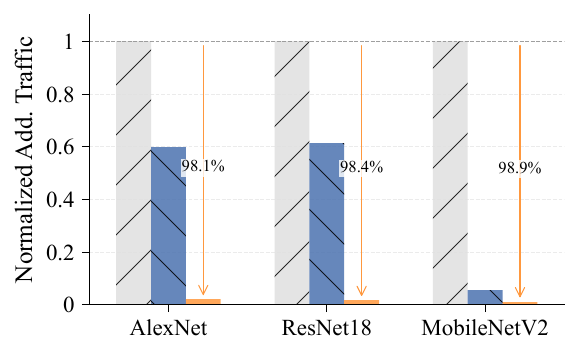}}
    \hfill
    \subfloat[Normalized energy overhead.\label{figures:evaluation_overall_energy}]{\includegraphics[width=0.32\linewidth]{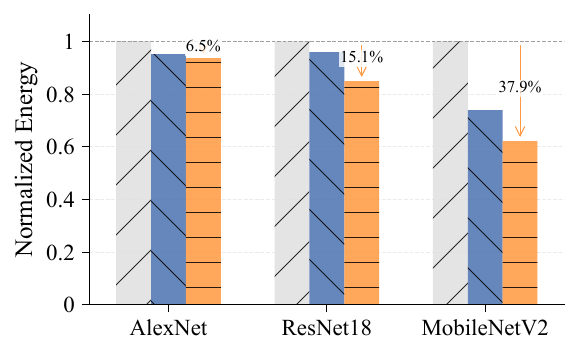}}
    \caption{Normalized latency, additional memory traffic, and energy overhead of the evaluated secure scheduling frameworks, with all metrics normalized to the Tile-Block-AC baseline.}
    \label{figures:evaluation_combined}
    \vspace{-10pt}
\end{figure*}

\subsubsection{Accelerator Configuration} \label{section:evaluation:exp:config}

Our experimental evaluation employs spatial DNN accelerator architectures with two-dimensional PE arrays and hierarchical memory organization. The architecture comprises PE-private register files for temporal data reuse, dedicated on-chip SRAM buffers for weights and feature maps, and off-chip DRAM for parameter storage. Following a unified security protocol, all model parameters and input activations are initially stored in unprotected external memory before being transferred to secure on-chip buffers during computation. The baseline architecture implements an Eyeriss-inspired row-stationary dataflow~\cite{chen2016eyeriss} that optimizes energy efficiency through spatial data reuse. The baseline PE array is configured as 14×12 processing elements (168 PEs total), with each PE containing private register files for localized data reuse. The on-chip memory hierarchy consists of three levels: PE-private registers, a global buffer (GLB) of 32KB shared across all PEs for intermediate feature map and weight buffering, and off-chip DRAM for large-scale model parameter and activation storage. We evaluate multiple architectural variants with different PE array dimensions (14×12, 28×24, and 42×36) and buffer sizes (16KB to 128KB) to assess scalability.

\subsubsection{Authenticated Cryptographic Engine} \label{section:evaluation:exp:crypto}

Our cryptographic engine employs the AES-GCM (Advanced Encryption Standard-Galois/Counter Mode) algorithm to simultaneously provide data confidentiality through AES encryption and integrity assurance via GHASH authentication. The hardware realization of AES-GCM~\cite{banerjee2017energy} supports three distinct architectural paradigms with different performance-resource trade-offs. Pipelined mode delivers maximum throughput, at the cost of increased area and power consumption. Serial mode processes encryption blocks sequentially, and achieves lower throughput but has minimal hardware resource requirements. Parallel mode employs multiple AES cores for concurrent block processing, and exhibits moderate throughput performance. Our implementation adopts a parallel AES-GCM architecture to achieve an efficient balance between performance and resource utilization. The AES module occupies 9.2 kGates, requires 11 cycles per block, and consumes 194.6 pJ per block. The GHASH module occupies 9.7 kGates, requires 8 cycles per block, and consumes 82.4 pJ per block. This architecture enables high-throughput processing of large data volumes while sustaining the low-latency operation required for secure DNN acceleration.

\subsubsection{Workloads and Simulation Configuration}

We evaluate these secure DNN accelerator frameworks using three representative workloads: AlexNet~\cite{krizhevsky2012imagenet}, ResNet18~\cite{he2016deep}, and MobileNetV2~\cite{sandler2018mobilenetv2}. These models span diverse architectural paradigms and computational characteristics essential for comprehensive security overhead analysis. AlexNet represents classical deep convolutional architectures with straightforward layer sequences, ResNet18 introduces residual connections that create complex inter-layer dependencies, while MobileNetV2 employs depthwise separable convolutions optimized for resource-constrained deployment scenarios. This selection enables systematic assessment of security mechanism impacts across varying network depths, architectural complexities, and computational patterns. Our experimental methodology leverages the synergistic integration of Timeloop~\cite{parashar2019timeloop} and Accelergy~\cite{accelergy} as complementary tools for comprehensive accelerator design space exploration. Timeloop serves as the architectural modeling engine, enabling detailed analysis of PE array configurations, memory hierarchies, and dataflow patterns under security constraints. Accelergy provides energy-aware evaluation capabilities, quantifying power consumption across different security implementations. This unified co-simulation addresses the intrinsic challenges of secure accelerator design through systematic co-optimization of computational patterns and hardware resource utilization, enabling exhaustive yet efficient exploration of security-performance trade-offs in deep learning acceleration.

\subsection{Overall Comparisons}

We compare three scheduling frameworks---Tile-Block-AC, Slice-Block-AC, and Chunk-Layer-AC---that represent different tradeoffs between authentication granularity and authentication-metadata overhead. Tile-Block-AC serves as the secure baseline, directly using a tile-based authentication strategy. Slice-Block-AC reduces some redundant security work through finer partitioning, but increases metadata overhead. Chunk-Layer-AC combines chunk-based authentication for feature maps with tile-based authentication for partial sums, targeting both redundancy and metadata cost. Furthermore, it implements a hierarchical authentication mechanism that optimizes security metadata management. This hierarchical design effectively reduces the off-chip memory storage and access overhead of authentication metadata. We next examine how these design choices affect latency, memory traffic, and energy.

\subsubsection{Latency Overhead}

\Cref{figures:evaluation_overall_latency}  presents the normalized latency overhead across three representative DNN workloads (AlexNet, ResNet18, and MobileNetV2), each normalized to Tile-Block-AC. Chunk-Layer-AC consistently outperforms both Slice-Block-AC and Tile-Block-AC across all benchmarks. For AlexNet, Chunk-Layer-AC reduces latency by $21.3\%$ and $24.1\%$ relative to Slice-Block-AC and Tile-Block-AC, respectively. The advantage becomes increasingly pronounced in deeper architectures: ResNet18 exhibits improvements of $48.6\%$ and $51.2\%$, while MobileNetV2 shows the most significant gains of $80.8\%$ and $87.1\%$.

These gains stem from two complementary effects. First, cross-layer chunking strategy removes redundant security work caused by overlap. Second, hierarchical authentication mechanism reduces unnecessary metadata movement by matching authentication granularity to data type. The former optimizes data reuse across adjacent layers, while the latter minimizes cryptographic overhead by tailoring authentication granularity to the specific computational characteristics of each layer. These synergistic optimizations enable Chunk-Layer-AC to achieve superior performance efficiency. The advantage is larger in networks with more complex layer interactions. In particular, ResNet18 benefits from cross-layer reuse around residual connections, significantly reducing cryptographic overhead. MobileNetV2's depthwise separable convolutions generate irregular memory access patterns. These patterns benefit substantially from Chunk-Layer-AC's adaptive scheduling, which dynamically adjusts authentication granularity based on computational characteristics. Therefore, Chunk-Layer-AC avoids the expensive tile-overlapping security computations that dominate conventional approaches.

\begin{figure}[!t]
    \centering
    \includegraphics[width=1\linewidth]{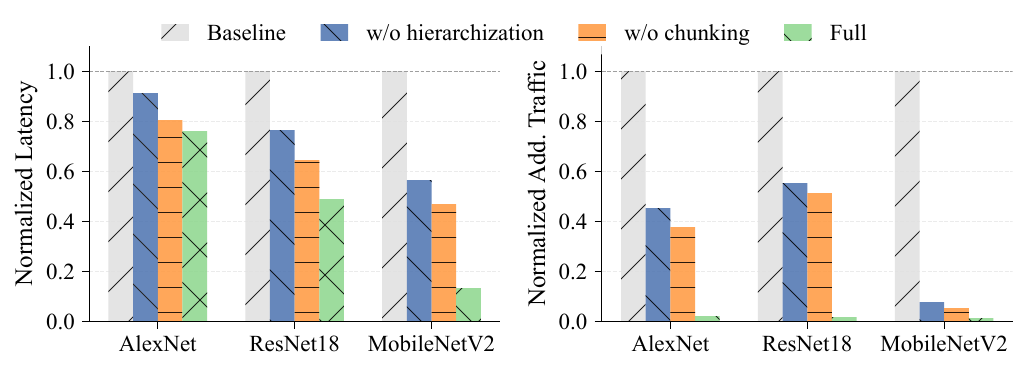}
    \caption{Ablation study results showing the impact of individual optimizations on normalized latency and additional memory traffic, with all metrics normalized to the Tile-Block-AC baseline.}
    \label{figures:evaluation_ablation}
    \vspace{-10pt}
\end{figure}

\subsubsection{Additional Memory Traffic}

\Cref{figures:evaluation_overall_traffic} presents normalized additional memory traffic relative to Tile-Block-AC, demonstrating the architectural advancements of {\design} in minimizing memory access overhead under security constraints. Chunk-Layer-AC reduces additional traffic by up to $97.4\%$ compared to Slice-Block-AC and by up to $98.9\%$ compared to Tile-Block-AC. The reduction comes from two sources: fewer redundant reads of overlapped data and fewer off-chip accesses for authentication metadata.

These performance gains originate from Chunk-Layer-AC's hierarchical authentication mechanism. It leverages a dual-granularity strategy---tile-based authentication for partial sums and chunk-based authentication for feature maps---to minimize off-chip accesses to security metadata. This approach significantly reduces the generation frequency of authentication metadata and relieves the memory bandwidth bottlenecks that typically restrict the performance of secure DNN accelerators. The effect is most pronounced in MobileNetV2, where Chunk-Layer-AC reduces additional traffic to only $1.1\%$ of Tile-Block-AC's overhead, compared with $5.3\%$ for Slice-Block-AC. The same trend appears across all three workloads, indicating that the benefit is not tied to a single model family.

\begin{figure*}[!t]
    \centering
    \includegraphics[width=0.99\linewidth]{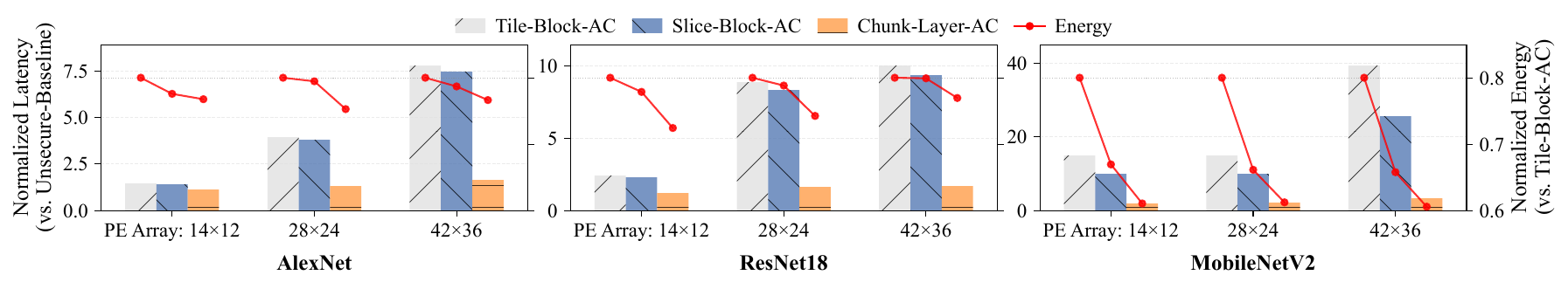}
    \caption{Normalized latency and energy consumption of secure scheduling frameworks across scaling PE array sizes.}
    \label{figures:evaluation_scale_pe}
    \vspace{-10pt}
\end{figure*}

\subsubsection{Energy Consumption} \label{section:evaluation:overall:energy}

\Cref{figures:evaluation_overall_energy} presents normalized energy consumption relative to Tile-Block-AC. The results reveal a clear performance hierarchy consistent with the latency and memory traffic findings. As shown in \Cref{figures:evaluation_overall_traffic}, once redundant data transfers and metadata accesses are removed, energy consumption falls accordingly. Chunk-Layer-AC reduces energy overhead by up to $15.9\%$ compared to Slice-Block-AC and by up to $37.9\%$ compared to Tile-Block-AC. The benefit is again larger for architecturally complex models (e.g., ResNet18 and MobileNetV2), where the baseline overheads are higher.

Even with increased chunk counts due to fragmentation, our hierarchical layer-level verification aggregates all chunk tags into a single layer-level tag via commutative XOR, writing back only one tag per layer to DRAM rather than one per chunk. Consequently, off-chip metadata traffic remains drastically lower than tile- and slice-based baselines, as confirmed by our experimental results, which show a $98.4\%$ and $98.9\%$ reduction in additional memory traffic for ResNet18 and MobileNetV2, respectively. Moreover, our evaluation on these residual-rich architectures demonstrates that Seculor achieves $51.2\%$ latency reduction over the tile baseline.

\subsubsection{Ablation Study} \label{section:evaluation:overall:ablation}

The ablation study in \Cref{figures:evaluation_ablation} isolates the contributions of chunk-based authentication and hierarchical verification against the Tile-Block-AC baseline. The full combination yields the lowest normalized latency across all workloads, with the largest gains on MobileNetV2 due to depthwise separable convolutions and irregular memory access that create substantial overlap-induced redundancy. Chunk-based authentication alone reduces latency moderately by eliminating redundant reads from tile-boundary misalignments, while hierarchical verification alone offers larger gains by aggregating per-chunk tags into a single layer-level tag, drastically cutting off-chip metadata traffic. Their combined effect is super-additive in deeper networks, confirming strong complementarity. Memory traffic follows the same trend: hierarchical verification contributes the bulk of the reduction by removing per-chunk tag reads, while chunking handles remaining redundant data fetches; together they nearly eliminate security-induced memory overhead.

\subsection{Architectural Scalability}

\Cref{figures:evaluation_scale_pe} evaluates normalized latency and energy overhead of the scheduling strategies across increasing PE array configurations ($14{\times}12$, $28{\times}24$, and $42{\times}36$). Chunk-Layer-AC consistently achieves the lowest latency and energy overhead across all PE configurations and workloads, confirming the efficacy of its cross-layer computation-security fusion strategy. In contrast, Tile-Block-AC and Slice-Block-AC exhibit substantial overhead growth as both PE array size and model complexity increase, demonstrating limited scalability under larger configurations.

\textbf{Latency Overhead.} The normalized latency results demonstrate that Chunk-Layer-AC scales much more gracefully than the baselines across all evaluated workloads. For AlexNet, Chunk-Layer-AC incurs a maximum overhead of only $1.61\times$, outperforming Slice-Block-AC ($7.45\times$) and Tile-Block-AC ($7.76\times$) at the $42{\times}36$ configuration. For ResNet18, Chunk-Layer-AC limits overhead to at most $1.69\times$, while Slice-Block-AC and Tile-Block-AC incur up to $9.35\times$ and $9.99\times$, respectively. The most compelling scalability advantage emerges in MobileNetV2, where Chunk-Layer-AC delivers latency reductions of up to $87.0\%$ and $91.6\%$ over Slice-Block-AC and Tile-Block-AC at the largest PE configuration, respectively. This trend indicates that the proposed scheduling strategy prevents security overhead from growing in proportion to compute parallelism.

\textbf{Energy Overhead.} The normalized energy results further validate Chunk-Layer-AC's efficiency advantage. Chunk-Layer-AC consistently achieves the lowest energy consumption across all configurations: it reduces energy overhead by up to $9.4\%$ for AlexNet, $15.1\%$ for ResNet18, and $38.9\%$ for MobileNetV2 compared to Tile-Block-AC. Slice-Block-AC also achieves moderate energy savings relative to Tile-Block-AC (up to $4.8\%$ for AlexNet, $4.3\%$ for ResNet18, and $28.5\%$ for MobileNetV2), but consistently falls short of Chunk-Layer-AC across all evaluated workloads and PE configurations. These results demonstrate that Chunk-Layer-AC's chunk-level scheduling strategy effectively minimizes redundant memory accesses and authentication overhead per inference, yielding superior energy efficiency that scales favorably with both model complexity and hardware resources.

\subsection{On-Chip Buffer Scalability}

\begin{figure}[!t]
    \centering
    \includegraphics[width=1\linewidth]{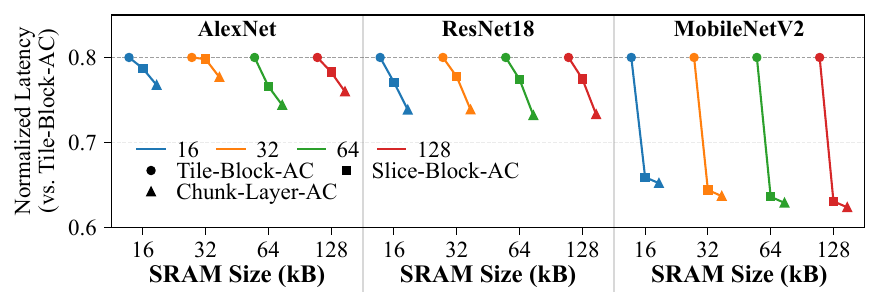}
    \caption{Normalized latency of secure scheduling frameworks across scaling SRAM buffer sizes, with results normalized to the Tile-Block-AC baseline.}
    \label{figures:evaluation_scale_buffer}
    \vspace{-10pt}
\end{figure}

\Cref{figures:evaluation_scale_buffer} characterizes the impact of on-chip buffer capacity on accelerator execution time across three representative DNN workloads, with buffer size scaled from 16\,kB to 128\,kB.

\textbf{Buffer Scaling Trend.} As buffer size increases from 16\,kB to 64\,kB, all three schedulers exhibit clear latency reductions, reflecting improved data reuse and reduced off-chip memory-access overhead enabled by the expanded on-chip storage. Chunk-Layer-AC benefits the most, achieving up to $15.8\%$ latency reduction for AlexNet and up to $23.8\%$ for ResNet18. Beyond 64\,kB, however, the marginal benefit diminishes markedly: the additional step from 64\,kB to 128\,kB yields less than $0.8\%$ further latency reduction across all workloads and schedulers. This saturation suggests that a 64\,kB buffer is sufficient to accommodate the dominant working sets of the evaluated architectures, and further increases in SRAM capacity provide negligible performance returns. From a hardware cost--performance standpoint, \textit{64\,kB therefore represents a practical design sweet spot}, delivering near-peak throughput while remaining within realistic area and power budgets.

\textbf{Cross-Method Comparison.} Chunk-Layer-AC consistently outperforms both other secure schedulers at every buffer configuration. The performance advantage is most pronounced at 64\,kB, where Chunk-Layer-AC achieves up to $11.2\%$ latency reduction over Tile-Block-AC for AlexNet, $13.6\%$ for ResNet18, and $34.3\%$ for MobileNetV2. Compared to Slice-Block-AC, Chunk-Layer-AC reduces latency by up to $4.8\%$ (AlexNet), $8.9\%$ (ResNet18), and $2.2\%$ (MobileNetV2) at 64\,kB. These results demonstrate that {\design}'s chunk-based authentication and cross-layer tiling optimization yield consistent and workload-agnostic performance gains regardless of on-chip buffer capacity. In summary, a 64\,kB on-chip SRAM provides the most cost-effective trade-off between hardware resource investment and execution efficiency, and is recommended as the practical design target for secure DNN accelerators.

\subsection{Sensitivity Analysis} \label{section:evaluation:sensitive}

Understanding the sensitivity of our results to memory bandwidth, cryptographic throughput, and metadata size is essential for delineating the practical deployment boundaries of our framework. We assess memory bandwidth sensitivity by varying DRAM bandwidth from the default 64 bytes/cycle to 32 bytes/cycle and 128 bytes/cycle, while the parallel AES-GCM engine operates at its full cryptographic throughput. Increasing bandwidth to 128 bytes/cycle yields negligible performance improvement, because the bottleneck shifts from the DRAM interface to the cryptographic engine. Conversely, reducing bandwidth to 32 bytes/cycle leads to moderate performance degradation, confirming that the default configuration represents a well-balanced design point. This trend aligns with findings in prior work such as SecureLoop~\cite{lee2023secureloop}. Metadata size exerts a secondary yet quantifiable effect: it scales the absolute metadata volume but does not alter the relative reduction of $91.2\%$, since our XOR-based aggregation reduces the number of tag accesses while the per-tag bit-width remains unchanged.

\section{Related Work} \label{section:related_work}

\noindent\textbf{Secure DNN Accelerators:} Existing secure DNN accelerators~\cite{zuo2021sealing, hua2022guardnn, lee2022tnpu, hua2022mgx} typically use block-based security units with MACs, which significantly increase memory-traffic overhead. Securator~\cite{shrivastava2023securator} introduces XOR-based aggregation that combines multiple MACs into one MAC stored in on-chip SRAM, reducing memory traffic but neglecting intra-layer and inter-layer overlap handling. In general, these methods do not adequately address the challenges posed by tile overlaps, leading to redundant computations and increased memory traffic.

\noindent\textbf{Scheduling Optimization for Security:} DNN layers require different tiling patterns to maximize resource utilization, potentially introducing authentication overhead and redundant computation. SecureLoop~\cite{lee2023secureloop} incorporates cryptographic overhead into scheduling and uses modular arithmetic and simulated annealing to optimize authentication block assignments. However, it treats tile overlaps as fixed constraints and only minimizes the resulting redundant traffic, without eliminating its root cause or distinguishing authentication policies for partial sums and feature maps. In contrast, {\design} eliminates overlap-induced redundancy through compile-time chunk-boundary analysis and employs hierarchical verification to reduce redundant computation and metadata traffic.

\noindent\textbf{On-Chip Security Management:} Recent work demonstrates that secure counters can be dynamically derived from computational patterns~\cite{hua2022mgx,hua2022guardnn,lee2022tnpu}, eliminating explicit counter storage and Merkle Tree structures. MGX~\cite{hua2022mgx} uses application-specific version numbering with on-chip status signals. TNPU~\cite{lee2022tnpu} adopts cache-centric version generation confined to on-chip memory. These works are orthogonal to our approach and can be combined to achieve more efficient secure DNN accelerator designs.
\section{Conclusion} \label{section:conclusion}

This work presented {\design}, a novel scheduling framework for secure DNN accelerators that reduces security overhead by leveraging the deterministic structure of inference workloads. To mitigate tile-induced overlap overhead, {\design} introduces a fusion-aware chunking scheme that eliminates redundant security operations by dividing overlapped tiles into chunk-based blocks, thereby reducing both redundant computations and memory traffic. In addition, {\design} uses a hierarchical authentication mechanism that applies tile-based authentication to partial sums and chunk-based authentication to feature maps, reducing authentication-metadata overhead. {\design} additionally adopts a modular design with a security scheduler that seamlessly integrates into existing DNN accelerators, enhancing flexibility and scalability. Experimental results show that {\design} outperforms prior secure scheduling approaches while preserving security guarantees. Crucially, these benefits are achieved entirely at the scheduling level, without modifying the underlying cryptographic primitives, making {\design} a practical and complementary add-on to existing secure accelerator infrastructures. These properties make {\design} a practical path toward secure DNN execution when both efficiency and deployability matter.

\section*{Acknowledgements} \label{section:acks}
We declare that the generative AI DeepSeek-V3 was used solely for limited editing and language refinement in this work.

\bibliographystyle{IEEEtran}
\bibliography{reference}

\vspace{-.4in}
\begin{IEEEbiography}[{\includegraphics[width=1.0in,height=1.26in,clip,keepaspectratio]{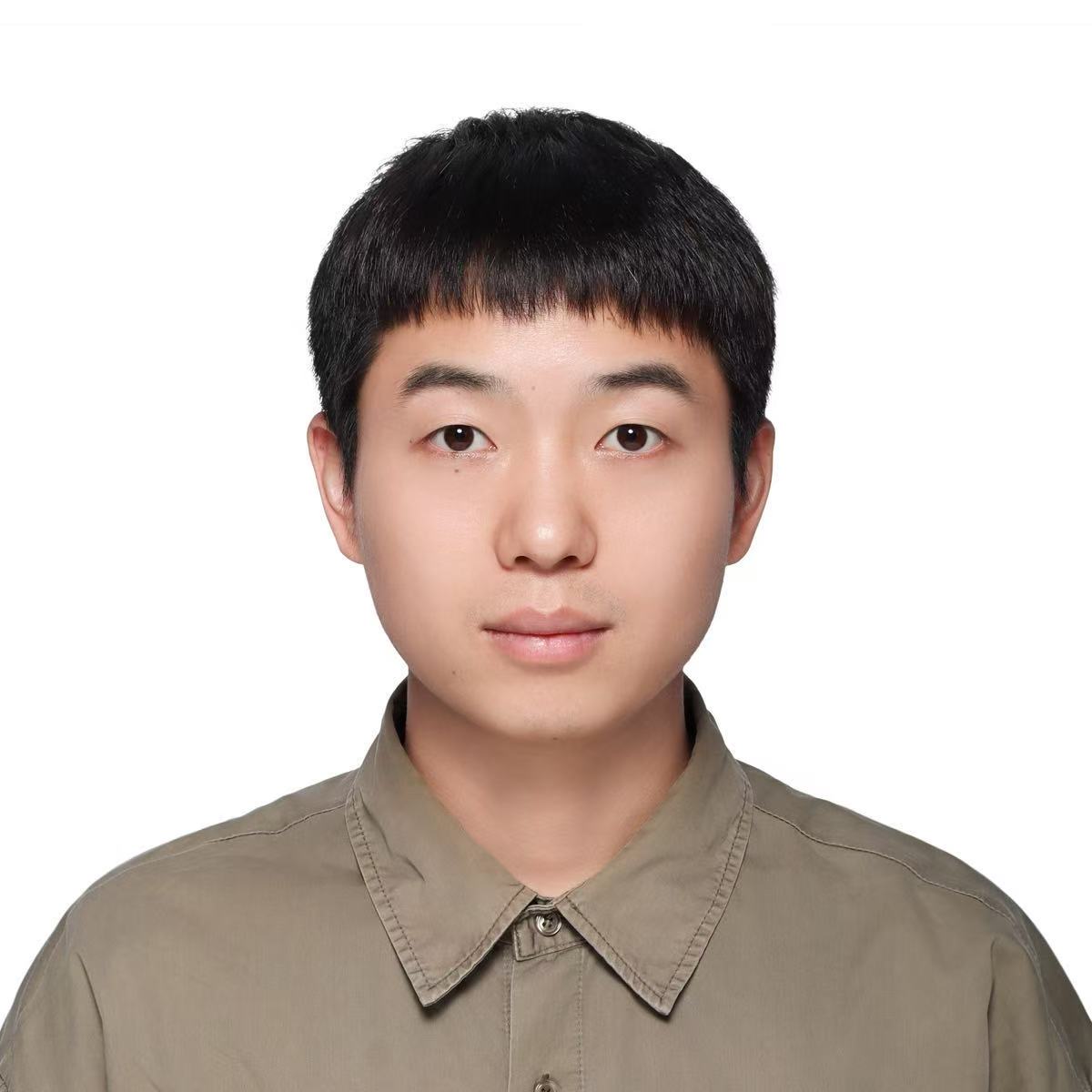}}]
  {Wei Xuan}
  is a tenure-track assistant professor at the Faculty of Computility Microelectronics, Shenzhen University of Advanced Technology. He was a postdoctoral fellow in the Department of Electronic and Computer Engineering at The Hong Kong University of Science and Technology and a member of the AI Chip Center for Emerging Smart Systems. He received his Ph.D. from the Institute of Computing Technology, Chinese Academy of Sciences. His research focuses on computer architecture, machine learning, and hardware security.
\end{IEEEbiography}

\vspace{-.4in}
\begin{IEEEbiography}
  [{\includegraphics[width=1.0in,height=1.26in,clip,keepaspectratio]{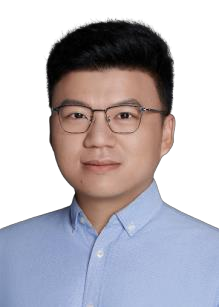}}]
  {Rongliang~Fu}
  received his Ph.D. in Computer Science and Engineering from The Chinese University of Hong Kong in January 2026, following an M.S. from the University of Chinese Academy of Sciences in June 2021 and a B.S. in Software Engineering from Northwestern Polytechnical University in June 2018. He has authored over 30 papers across major journals (IEEE TC and IEEE TCAD) and conferences(DAC, DATE, ICCAD, etc.). His research spans electronic design automation and EDA for superconducting electronics.
\end{IEEEbiography}

\vspace{-.4in}
\begin{IEEEbiography}[{\includegraphics[width=1.0in,height=1.26in,clip,keepaspectratio]{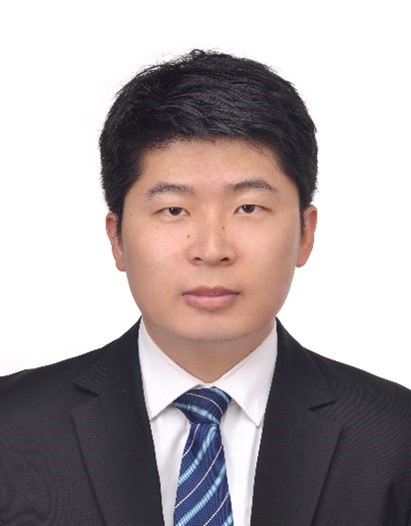}}]
  {Zihao Xuan}
  received the B.S. degree in mechano-electronic engineering from Xidian University, Xi’an, China, in 2017, and the Ph.D. Degree in electronic science and technology from University of Science and Technology of China, Hefei, Anhui, China, in 2023.
  He is currently a Postdoctoral fellow in the Department of Electronic and Computer Engineering at The Hong Kong University of Science and Technology and a member of the AI Chip Center for Emerging Smart Systems (ACCESS) under InnoHK. His current research areas include efficient neural network algorithms and their implementation on chips.
\end{IEEEbiography}

\vspace{-.4in}
\begin{IEEEbiography}[{\includegraphics[width=1.0in,height=1.26in,clip,keepaspectratio]{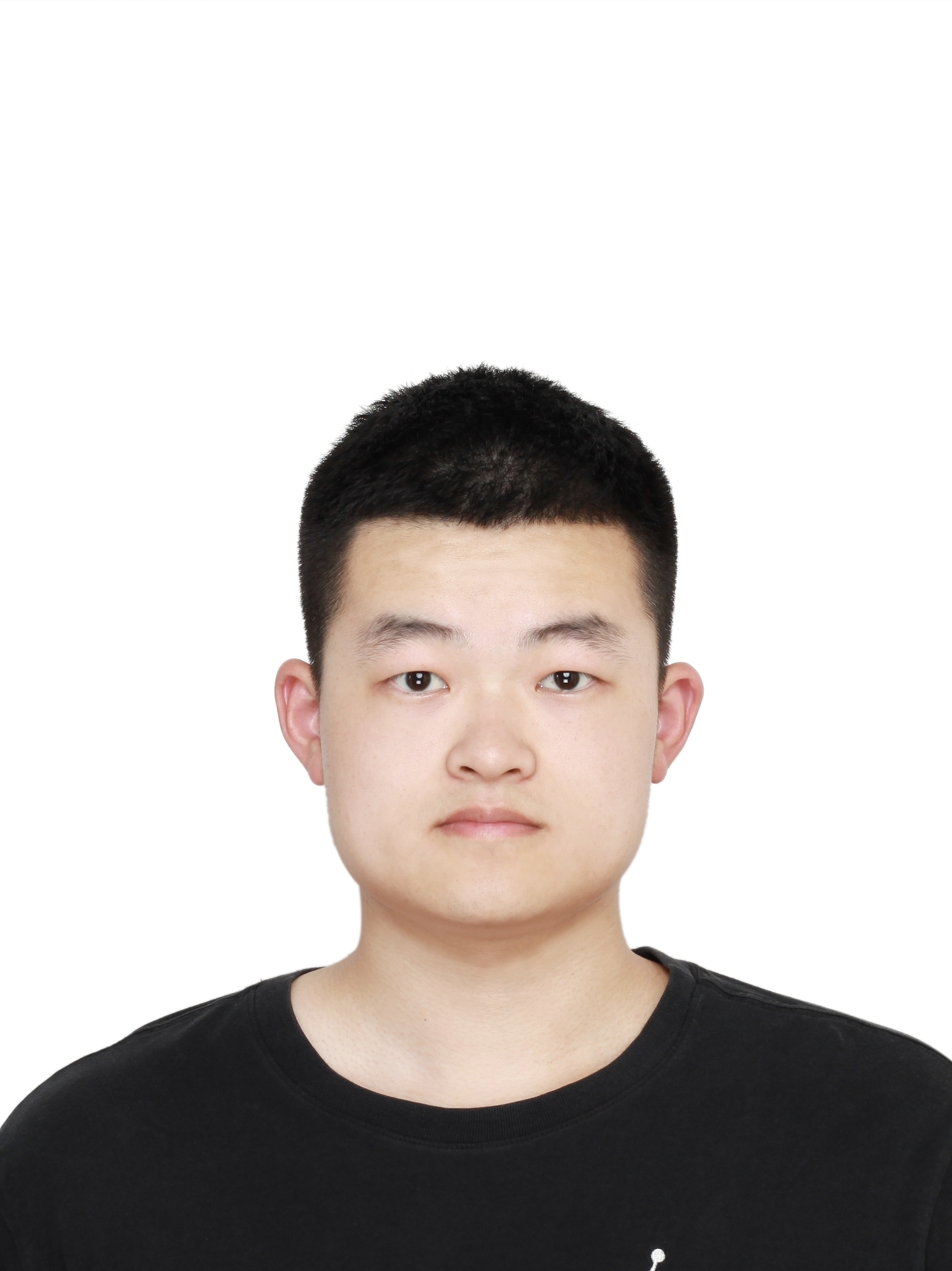}}]
  {Li Bo}
  received his B.S. degree in communication engineering from Beijing University of Posts and Telecommunications, Beijing, China, in 2021 and his MS degree in computer technology from the University of Chinese Academy of Sciences, Beijing, China, in 2024. He is currently a Ph.D. student at the Department of Computer Science and Engineering, The Chinese University of Hong Kong. His research interests include electronic design automation and computer architecture.
\end{IEEEbiography}

\vspace{-.4in}
\begin{IEEEbiography}[{\includegraphics[width=1.0in,height=1.26in,clip,keepaspectratio]{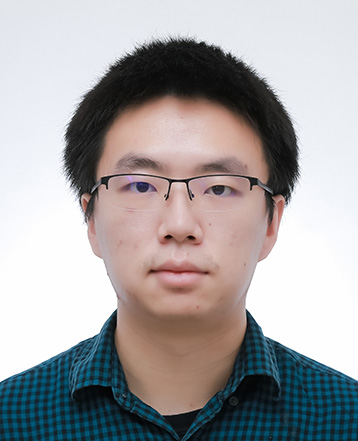}}]
  {Lang Feng}
  received his Ph.D. degree in computer engineering from Texas A\&M University in 2020, and his B.E. degree in electronic science and technology (microelectronic technology) from University of Electronic Science and Technology of China in 2016. He is currently an associate professor at the School of Integrated Circuits of Sun Yat-sen University Shenzhen Campus. Previously, he worked for Nanjing University. His research interests include electronic design automation, computer architecture, etc.
\end{IEEEbiography}

\vspace{-.4in}
\begin{IEEEbiography}[{\includegraphics[width=1.0in,height=1.26in,clip,keepaspectratio]{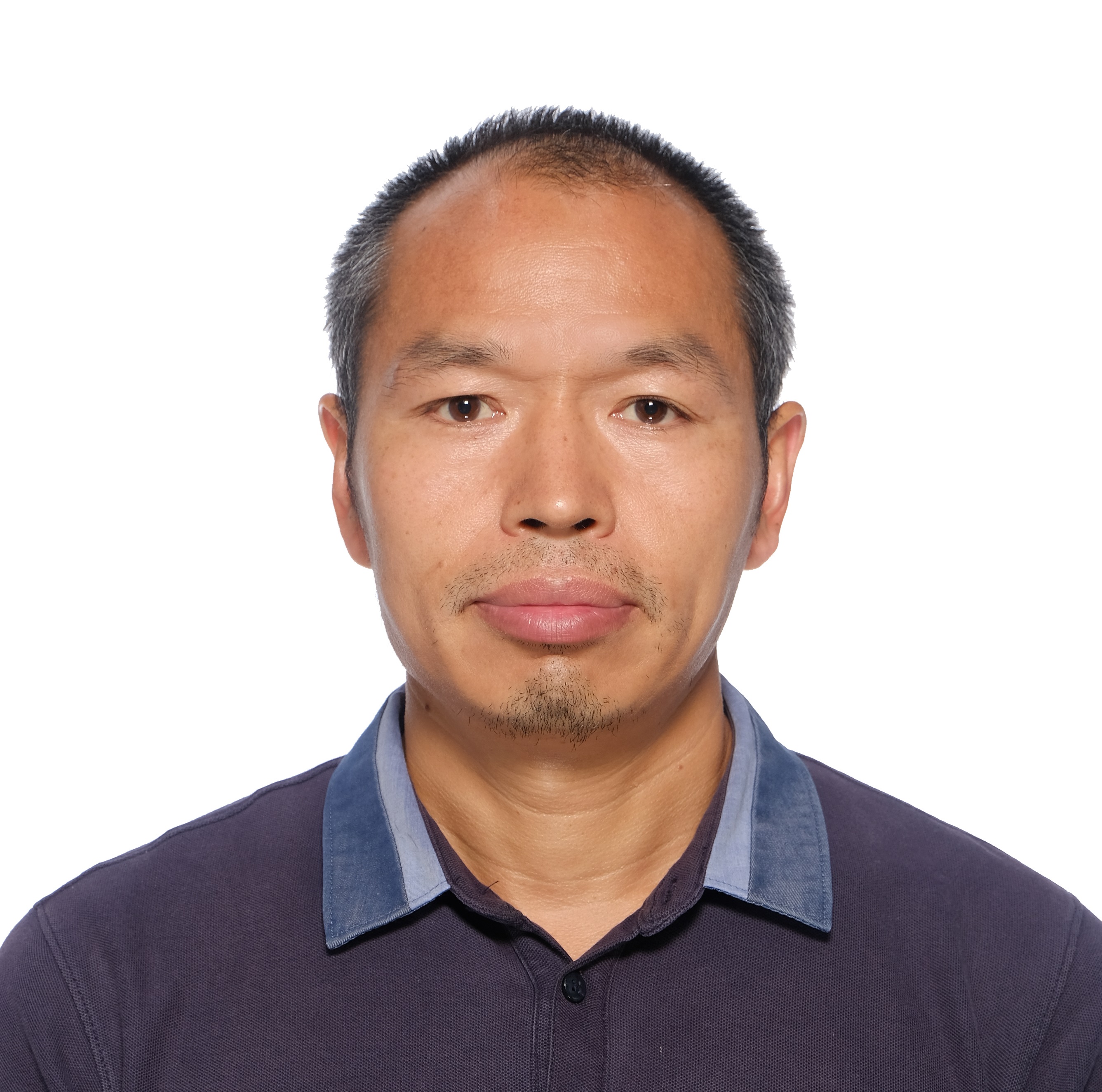}}]
  {Yuzhong Jiao}
  received the M.S. degree in Measurement Technologies and Instruments from Xiamen University, Xiamen, China in 2001, and the Ph.D degree in Microelectronics and Solid-state Electronics from Peking University, Beijing, China, in 2009. 
  He is currently a Principal Researcher in AI Chip Center for Emerging Smart Systems (ACCESS), Hong Kong, China. His current research interests include research and development on AI accelerators, interconnection, and advanced packaging.
\end{IEEEbiography}

\vspace{-.4in}
\begin{IEEEbiography}[{\includegraphics[width=1.0in,height=1.26in,clip,keepaspectratio]{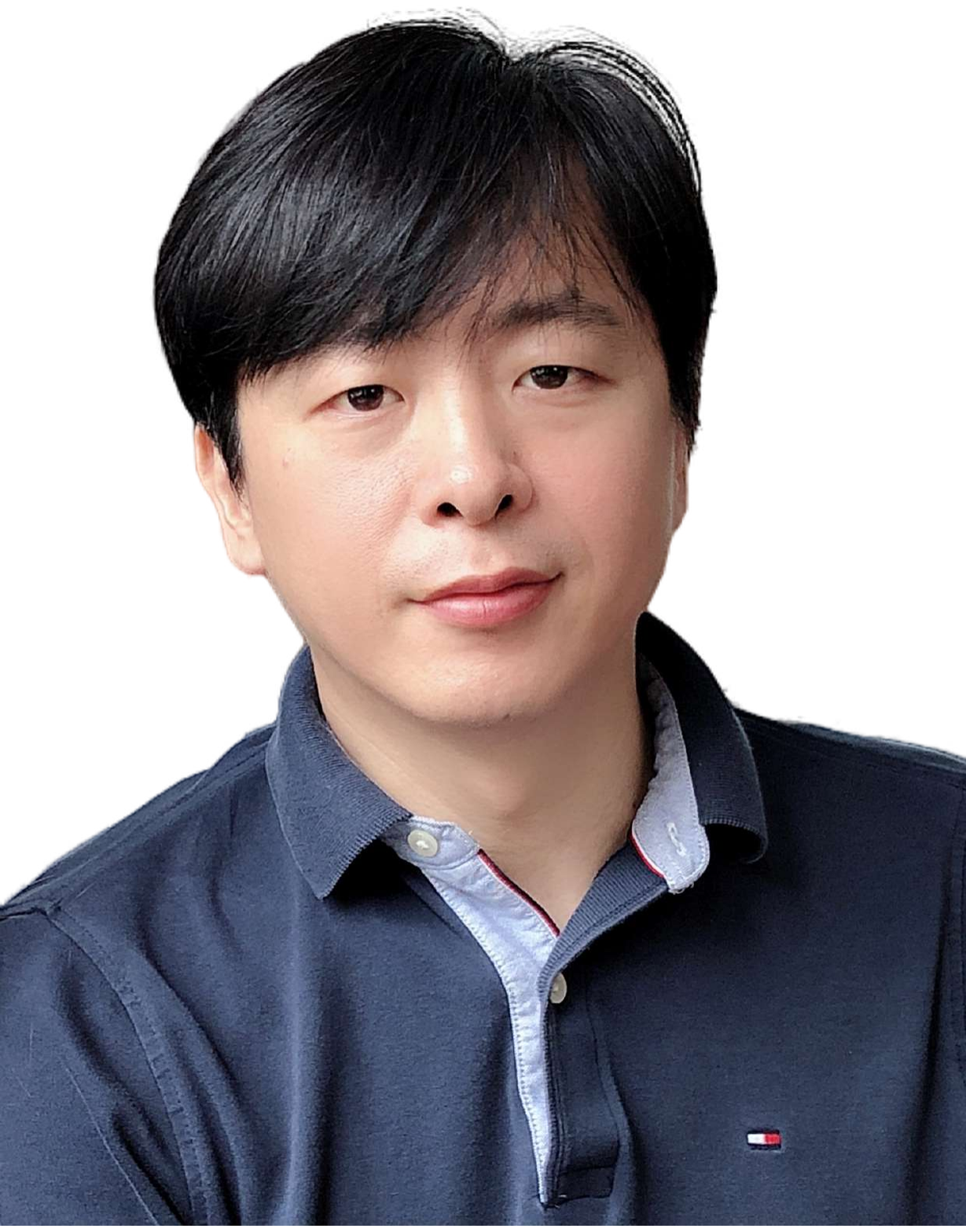}}]
  {Tsung-Yi Ho}
  (F'24)
  is a Professor in the Department of Computer Science and Engineering, The Chinese University of Hong Kong (CUHK). He received his Ph.D. in Electrical Engineering from National Taiwan University in 2005. His research interests include several areas of computing and emerging technologies, especially in the design automation of microfluidic biochips. He was a recipient of the Best Paper Award at the IEEE Transactions on Computer-Aided Design of Integrated Circuits and Systems in 2015. Currently, he serves as the VP Conferences of IEEE CEDA, and the Executive Committee of ASP-DAC and ICCAD. He is a Distinguished Member of ACM and a Fellow of IEEE.
\end{IEEEbiography}

\vspace{-.4in}
\begin{IEEEbiography}[{\includegraphics[width=1.0in,height=1.26in,clip,keepaspectratio]{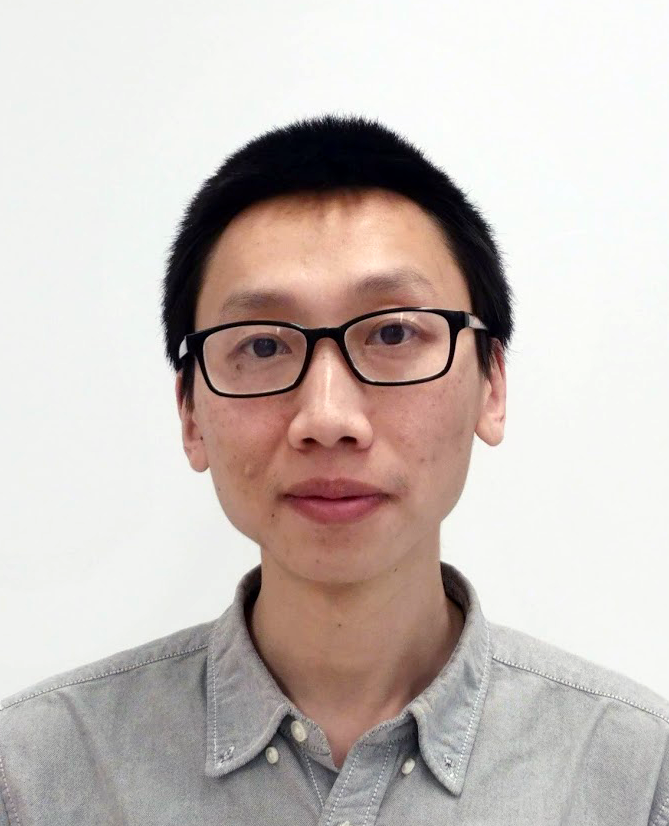}}]
  {Zhongrui Wang}
  is a tenured associate professor at the School of Microelectronics at Southern University of Science and Technology, a awardee of the NSFC Excellent Youth Fund (Hong Kong and Macau), and a Clarivate Highly Cited Researcher. Prior to joining SUSTech, he was an assistant professor in the Department of Electrical and Electronic Engineering at the University of Hong Kong. He earned his Bachelor's degree (First Class Honors) and Ph.D. from Nanyang Technological University in Singapore. Dr. Wang's research primarily focuses on novel AI hardware and algorithm co-design.
\end{IEEEbiography}

\end{document}